%% file: paper2.tex
\documentclass[a4paper,fleqn,usenatbib]{mnras}

\usepackage{newtxtext}
\usepackage{newtxmath}

\usepackage[T1]{fontenc}
\usepackage{ae,aecompl}

\usepackage{graphicx}	%

\ifdefined\Bbbk{}
    \let\Bbbk\relax
\else
\fi
\ifdefined\bigtimes{}
    \let\bigtimes\relax
\else
\fi
\usepackage[separate-uncertainty=true,multi-part-units=single,alsoload=astro]{siunitx}
\usepackage{pgf}
\usepackage{xprintlen}
\usepackage{booktabs}
\usepackage{mathabx}
\usepackage{wasysym}

\usepackage{orcidlink}

\usepackage{tikz}
\usetikzlibrary{shapes.geometric}

\hypersetup{pdfauthor={D. Evensberget et al.}}  %

\newcommand{\batsrus}{{\textsc{bats-r-us}}}
\newcommand{\swmf}{{\textsc{swmf}}}
\newcommand{\awsom}{{\textsc{awsom}}}
\renewcommand{\pluto}{{\textsc{pluto}}} %
\newcommand{\toupies}{TOUPIES}

\newcommand{\SF}{{\Phi}}  %
\newcommand{\ymaxmingeom}{y_\text{max}/y_\text{min}} %
\newcommand{\ymaxminstat}{y_{0.975}/y_{.025}} %

\definecolor{darkgreen}{rgb}{0.0, 0.5, 0.0}

\newcommand{\Wind}{\textsc{w}} %
\newcommand{\Alfven}{\textsc{a}} %
\newcommand{\Star}{\text{\bigstar}} %

\renewcommand{\Earth}{{\mathchoice{}{}{\scriptscriptstyle}{}\oplus}} %
\renewcommand{\Sun}{{\mathchoice{}{}{\scriptscriptstyle}{}\odot}} %
\renewcommand{\Star}{{\mathchoice{}{}{\scriptscriptstyle}{}\bigstar}} %

\defcitealias{2016MNRAS.457..580F}{F16}
\defcitealias{2018MNRAS.474.4956F}{F18}
\defcitealias{paper1}{Paper I}

\newcommand{\ZDI}{{\mathchoice{}{}{\scriptscriptstyle}{}\mathrm{ZDI}}}

\DeclareSIUnit\year{yr}
\DeclareSIUnit\astronomicalunit{au}
\DeclareSIUnit\parsec{pc}
\DeclareSIUnit\erg{erg}
\DeclareSIUnit\gauss{G}
\DeclareSIUnit\mSun{\mbox{\(M_\Sun\)}}
\DeclareSIUnit\rSun{\mbox{\(R_\Sun\)}}
\DeclareSIUnit\mEarth{\mbox{\(M_\Earth\)}}
\DeclareSIUnit\rEarth{\mbox{\(R_\Earth\)}}

\renewcommand{\vec}[1]{\boldsymbol{#1}} %
\newcommand{\uvec}[1]{\boldsymbol{\hat{#1}}} %

\newcommand\thefont{\expandafter\string\the\font} %

\renewcommand{\propto}{\mathrel{\mathchar"939}} %
\newcommand{\mysim}{{\sim}} %

\newcommand{\Unscaled}[2] %
{\ignorespaces  %
    \tikz[baseline={([yshift=-.75ex]current bounding box.center)}]
    {
        \node
        [
            circle,
            anchor=base,
            text=white,
            fill=#2,
            inner sep=0.7pt
        ]
        (M)
        {
            \tiny{#1}
        };
    }~\ignorespaces  %
}
\newcommand{\Scaled}[2] %
{\ignorespaces  %
    \tikz[baseline={([yshift=-.75ex]current bounding box.center)}]
    {
        \node
        [
            star,
            anchor=base,
            star point ratio=2,
            text=white,
            fill=#2,
            inner sep=0.0pt
        ](M)
        {
            \tiny{#1}
        };
    }~\ignorespaces  %
}

\definecolor{C0}{HTML}{808080}  %
\definecolor{C1}{HTML}{ff7f0e}
\definecolor{C2}{HTML}{2ca02c}

\title[Winds of Coma and Her-Lyr]{The winds of young, Solar-type stars in Coma Berenices and Hercules-Lyra}
\author[D. Evensberget et al.]{D. Evensberget \orcidlink{0000-0001-7810-8028}\(^{1}\)\thanks{E-mail: dag.evensberget@usq.edu.au (USQ)},
    B. D. Carter \orcidlink{0000-0003-0035-8769}\(^{1}\),
    S. C. Marsden \orcidlink{0000-0001-5522-8887}\(^{1}\),
    L. Brookshaw \orcidlink{0000-0001-8503-0062}\(^{1}\), 
    \newauthor
    C. P. Folsom \orcidlink{0000-0002-9023-7890}\(^{2}\),
    and 
    R. Salmeron \orcidlink{0000-0002-1956-4493}\(^{1}\)
    \\
    \(^{1}\)Centre for Astrophysics, University of Southern Queensland, Toowoomba, Queensland 4350, Australia\\
    \(^{2}\)Tartu Observatory, University of Tartu, Observatooriumi 1, T\~{o}ravere, 61602 Tartumaa, Estonia
}

\date{Accepted XXX.\@ Received YYY;\@ in original form ZZZ}
\pubyear{2021}
  
\begin{document}
\label{firstpage}
\pagerange{\pageref{firstpage}--\pageref{lastpage}}
\maketitle
\begin{abstract}
    We present wind models of ten young Solar-type stars in the Hercules-Lyra association and the Coma Berenices cluster  aged around 
    \SI{\mysim 0.26}{\giga\year}
    and 
    \SI{\mysim 0.58}{\giga\year}
    respectively. Combined with five previously modelled stars in the Hyades cluster, 
    aged 
    \SI{\mysim 0.63}{\giga\year}, 
    we obtain a large atlas of fifteen observationally based wind models. We find varied geometries, multi-armed structures in the equatorial plane, and a greater spread in quantities such as the angular momentum loss. In our models we infer variation of a factor of \(\mysim 6\) in wind angular momentum loss \(\dot J\) and a factor of \(\mysim 2\) in wind mass loss \(\dot M\) based on magnetic field geometry differences when adjusting for the unsigned surface magnetic flux.
    We observe a large variation factor of \(\mysim 4\) in wind pressure for an Earth-like planet; we attribute this to variations in the `magnetic inclination' of the magnetic dipole axis with respect to the stellar axis of rotation.
    Within our models, we observe a tight correlation between unsigned open magnetic flux and angular momentum loss. 
    To account for possible underreporting of the observed magnetic field strength we investigate a second series of wind models where the magnetic field has been scaled by a factor of 5. This gives \(\dot M\propto B^{0.4}\) and \(\dot J\propto B^{1.0}\) as a result of pure magnetic scaling.
\end{abstract}


\input{paper2-body}

\bibliographystyle{mnras}
\bibliography{bibliography} %

\appendix
\input{pooled-series}

\bsp	%
\label{lastpage} %
\end{document}

%% file: paper2-body.tex
\section{Introduction}\label{sec:intro}
\begin{table*}
    \caption[]{
        Fundamental parameters of the ten stars modelled in this study from \citetalias{2016MNRAS.457..580F}, \citetalias{2018MNRAS.474.4956F} and references therein. The stellar period of rotation \(P_\text{rot}\), radius \(R\) and mass \(M\) are used in the 
        magnetohydrodynamic simulations, along with the associated surface magnetic maps in Fig.~\ref{fig:magnetograms-zdi}. 
        In the radius and mass columns the values are scaled to the Solar values of mass \(M_\Sun=\SI{1.99e30}{\kilogram}\) and radius \(R_\Sun=\SI{6.96e8}{\meter}\). 
        In this paper we abbreviate the full star names to the simulation case names given in the first column of this table. To aid the reader, each case name is prepended by a unique identifier symbol used throughout this paper.
    }\label{tab:observed_quantities}
    \centering
    \begin{tabular}{lllcccccl}
        \toprule
        Case name & Full name
        \citepalias[see][]{2016MNRAS.457..580F,2018MNRAS.474.4956F}
         & Association & Type & \(P_\text{rot}\) &  Age    &  \(R\)         & \(M\)     & Reference \\
                &           &        &      & (\si{\day})              &  (\si{\mega\year})  &  \((R_\Sun)\) & \((M_\Sun)\)      \\
        \midrule
        \Unscaled{5}{C1} AV  523 & \href{http://simbad.u-strasbg.fr/simbad/sim-id?Ident=Cl*+Melotte+111+AV+523  }{Cl* Melotte 111 AV 523 } & Coma Berenices &  K2  & \(11.10 \pm 0.20\) & \(584 \pm  10\)& \(0.72 \pm 0.033\)& \(0.80 \pm 0.05\) & \citetalias{2018MNRAS.474.4956F}\\ 
        \Unscaled{6}{C1} AV 1693 & \href{http://simbad.u-strasbg.fr/simbad/sim-id?Ident=Cl*+Melotte+111+AV+1693 }{Cl* Melotte 111 AV 1693} & Coma Berenices &  G8  & \(9.05 \pm 0.10\)  & \(584 \pm  10\)& \(0.83 \pm 0.030\)& \(0.90 \pm 0.05\) & \citetalias{2018MNRAS.474.4956F}\\ 
        \Unscaled{7}{C1} AV 1826 & \href{http://simbad.u-strasbg.fr/simbad/sim-id?Ident=Cl*+Melotte+111+AV+1826 }{Cl* Melotte 111 AV 1826} & Coma Berenices &  G9  & \(9.34 \pm 0.15\)  & \(584 \pm  10\)& \(0.80 \pm 0.041\)& \(0.85 \pm 0.04\) & \citetalias{2018MNRAS.474.4956F}\\ 
        \Unscaled{8}{C1} AV 2177 & \href{http://simbad.u-strasbg.fr/simbad/sim-id?Ident=Cl*+Melotte+111+AV+2177 }{Cl* Melotte 111 AV 2177} & Coma Berenices &  G6  & \(8.98 \pm 0.12\)  & \(584 \pm  10\)& \(0.78 \pm 0.033\)& \(0.90 \pm 0.04\) & \citetalias{2018MNRAS.474.4956F}\\ 
        \Unscaled{9}{C1} TYC 1987 & \href{http://simbad.u-strasbg.fr/simbad/sim-id?Ident=TYC+1987-509-1          }{TYC 1987-509-1        } & Coma Berenices &  G7  & \(9.43 \pm 0.10\)  & \(584 \pm  10\)& \(0.83 \pm 0.033\)& \(0.90 \pm 0.05\) & \citetalias{2018MNRAS.474.4956F}\\ 
        \midrule
        \Unscaled{A}{C2} DX Leo   & \href{http://simbad.u-strasbg.fr/simbad/sim-id?Ident=DX+Leo                }{DX Leo                  }  & Hercules-Lyra  &  G9  & \(5.38 \pm 0.07\)  & \(257 \pm  46\)& \(0.81 \pm 0.026\)& \(0.90 \pm 0.04\) & \citetalias{2016MNRAS.457..580F}\\ 
        \Unscaled{B}{C2} EP Eri   & \href{http://simbad.u-strasbg.fr/simbad/sim-id?Ident=EP+Eri                }{EP Eri                  }  & Hercules-Lyra  &  K1  & \(6.76 \pm 0.20\)  & \(257 \pm  46\)& \(0.72 \pm 0.081\)& \(0.85 \pm 0.05\) & \citetalias{2018MNRAS.474.4956F}\\ 
        \Unscaled{C}{C2} HH Leo   & \href{http://simbad.u-strasbg.fr/simbad/sim-id?Ident=HH+Leo                }{HH Leo                  }  & Hercules-Lyra  &  G8  & \(5.92 \pm 0.02\)  & \(257 \pm  46\)& \(0.84 \pm 0.030\)& \(0.95 \pm 0.05\) & \citetalias{2018MNRAS.474.4956F}\\ 
        \Unscaled{D}{C2} V439 And & \href{http://simbad.u-strasbg.fr/simbad/sim-id?Ident=V439+And              }{V439 And                }  & Hercules-Lyra  &  K0  & \(6.23 \pm 0.01\)  & \(257 \pm  46\)& \(0.92 \pm 0.099\)& \(0.95 \pm 0.05\) & \citetalias{2016MNRAS.457..580F}\\ 
        \Unscaled{E}{C2} V447 Lac & \href{http://simbad.u-strasbg.fr/simbad/sim-id?Ident=V447+Lac              }{V447 Lac                }  & Hercules-Lyra  &  K1  & \(4.43 \pm 0.05\)  & \(257 \pm  46\)& \(0.81 \pm 0.089\)& \(0.90 \pm 0.04\) & \citetalias{2016MNRAS.457..580F}\\ 
        \bottomrule
    \end{tabular}     
\end{table*}
The age span linking the end of the stellar contractive phase and the onset of the 
\citet{1972ApJ...171..565S} relationship \(P_\text{rot}\propto t^{1/2}\) between stellar period of rotation and stellar age extends 
from approximately \SIrange{.1}{.6}{\giga\year} for Solar-type stars \citep{2013A&A...556A..36G,2015A&A...577A..98G}.
Stars enter this `pre-Skumanich spin-down phase' with a wide range of rotation periods \citep{1993AJ....106..372E} depending on the specifics of the preceding contractive phase. For the Skumanich relationship to take hold at \SI{.6}{\giga\year}, rapidly rotating stars must shed angular momentum more efficiently than slowly rotating stars.

\citet{2003ApJ...586..464B} found a bimodal period distribution of `fast and slow rotators' in the pre-Skumanich spin-down phase and suggested that the bimodal distribution of rotation periods arise as the `magnetically saturated' fast rotators are unable to effectively shed angular momentum \citep{1991ApJ...376..204M} by the wind mechanisms that apply in the slow rotator group. 
Many models of stellar spin-down \citep{1988ApJ...333..236K,1991ASIC..340...41B,1995ApJ...441..865C} thus invoke a threshold rotation velocity and/or magnetic field above which angular momentum shedding is inhibited in order to permit spin braking laws to reproduce the Skumanich relationship past \SI{.6}{\giga\year}. Depending on the choice of model the threshold rotation velocity can be 3--15 times the Solar rotational velocity~\citep{2015ApJ...799L..23M,2016A&A...587A.105A}.

The dominant mechanism of stellar spin-down is angular momentum shedding by means of the 
coupling between the stellar wind and the magnetic field \citep{1962AnAp...25...18S,1967ApJ...148..217W} and the resulting magnetic and dynamic torque components acting on the star. Thus
mathematical descriptions of the stellar wind angular momentum loss \(\dot J\) and its history must reproduce the Skumanich relationship from a wide range of initial periods of rotation in order to agree with observations.
The effect of the star's magnetic field on the wind angular momentum loss is well known:
the angular momentum loss is increased as if the magnetic field holds the escaping wind matter in co-rotation with the star until the wind speed exceeds the Alfvén wave speed \(u_\Alfven\) \citep{1942Natur.150..405A}, thus greatly increasing \(\dot J\). 
The Alfvén radius \(R_\Alfven\) at which the wind speed exceeds \(u_\Alfven\) varies with the magnetic field strength, wind density and wind velocity. 

The importance of the Alfvén radius and the magnetic field geometry is seen in one-dimensional and two-dimensional solar and stellar wind models \citep{1967ApJ...148..217W,1968MNRAS.138..359M,1984LNP...193...49M,1988ApJ...333..236K} in forms such as \(\dot J \propto P_\text{rot}^{-1} \dot M R_\Alfven^n\), where \(n\) is magnetic geometry parameter. The dependence of these models on the wind mass loss \(\dot M\) can limit their usefulness as \(\dot M\) is itself notoriously difficult to constrain observationally, and there is much uncertainty about the behaviour of \(\dot M\) for stars younger than \SI{\mysim 0.6}{\giga\year} \citep{2005ApJ...628L.143W,2014ApJ...781L..33W}. 

Numerical simulations permit the simultaneous reconstruction of \(\dot M\) and \(\dot J\) by solving the magnetohydrodynamic (MHD) equations; this, however, requires 
a model of the stellar surface magnetic field, 
and setting a coronal temperature 
    \citep[e.g.\ ][]{2009PhDT........94V,2014ApJ...783...55C,2018MNRAS.476.2465O}
and/or prescribing a model of coronal heating \citep{2014ApJ...782...81V}.
Depending on the models used, other parameters may need to be estimated; these may include the magnetic filling factor \citep{2011SSRv..158..339S}, the Poynting flux \citep{2020A&A...635A.178B}, the wave turbulence \citep{2011ApJ...743..197C,2015RSPTA.37340148C}, as well as other parameters.

As the stellar rotational energy ultimately sustains stellar magnetic fields, a greater variation in the relation between age, period of rotation, stellar magnetic fields, and stellar winds may be expected in the pre-Skumanich spin-down phase, with its range of rotation periods
and differing states of magnetic saturation, than in later phases of a star's lifespan. 
Depending on the saturation mechanism, on could also expect a greater variation in the relation between the surface magnetic field strength and wind mass- and angular momentum loss, for example if there is an physical increase in field complexity for more rapidly rotating stars \citep{2018ApJ...862...90G}.

In this work we study the pre-Skumanich spin-down phase stellar winds by creating wind models of young, Solar-type stars in the Coma Berenices cluster and the Hercules-Lyra association with well characterised ages of \SI{584\pm10}{\mega\year} and \SI{257\pm46}{\mega\year} respectively \citep{2018MNRAS.474.4956F}; these stars are in the late and middle pre-Skumanich spin-down phase. 
The Coma Berenices stars in our sample are slow rotators for their age, at and below the 25\textsuperscript{th} percentile in the fast\nobreakdash-, medium\nobreakdash-, and slow rotator classification of \citet{2013A&A...556A..36G}. The Hercules-Lyra stars of our sample exhibit some more variation and would be classified as slow to medium rotators for their age, sitting mostly below the 50\textsuperscript{th} percentile. Depending on the saturation threshold angular velocity \(\Omega = 2\pi/P_\text{rot}\), some of the sample stars may be in the unsaturated regime where the shedding of angular momentum is inhibited.

The wind models presented here are based on surface magnetic maps by \citet{2016MNRAS.457..580F,2018MNRAS.474.4956F}, hereafter \citetalias{2016MNRAS.457..580F,2018MNRAS.474.4956F}.
The lack of rotational symmetry of the stars' surface magnetic fields, in particular the offset between the magnetic dipole axis and the stellar rotation axis mandate the use of three-dimensional numerical simulations. 
We use the \citetalias{2016MNRAS.457..580F,2018MNRAS.474.4956F} magnetic maps to drive the numerical Alfvén wave Solar model 
\citep[\awsom{}, ][]{2013ApJ...764...23S, 2014ApJ...782...81V},
which is part of the Space weather modelling framework  \citep[\swmf{},][]{1999JCoPh.154..284P,2012JCoPh.231..870T}. 
In the \awsom{} model the corona is heated by Alfvén waves, which are thought to
originate in the stellar interior.
With its inner boundary in the chromosphere, \awsom{} incorporates the transition region, corona, and inner astrosphere; an Alfvén wave energy flux is prescribed at the model's inner boundary. In addition to the regular magnetohydrodynamical quantities, the wind maps give the electron-, ion-, and Alfvén wave pressure at each point in the solution. 
From the wind maps we calculate the steady-state wind mass loss and angular momentum loss, the wind pressure and spatial variation of the wind pressure for an Earth-like planet\footnote{In the context of this paper, an `Earth-like planet' means that we use the orbital elements of the Earth along with the Earth's radius and current-day dipolar magnetic field strength.}, and other relevant quantities in a self-consistent manner. 

By combining the wind models presented in this work with the wind models of young Solar-type stars 
from our previous work \citet{paper1} on the \SI{625\pm50}{\mega\year} old Hyades cluster, 
hereafter \citetalias{paper1},
we obtain an `atlas' of stellar wind models from which we are able to formulate scaling relations between aggregate quantities and the surface magnetic field strength and surface magnetic flux. 

The rest of this paper is outlined as follows: 
In Section~\ref{sec:Observations} we describe the surface magnetic maps and how they are obtained using Zeeman-Doppler imaging; 
in Section~\ref{sec:Simulations} we describe the model equations and numerical model;
in Section~\ref{sec:results} we give an overview of our model results including aggregate quantities calculated from the wind models such as mass loss; 
in Section~\ref{sec:Discussion} we examine trends in aggregate quantities within our own dataset;
in Section~\ref{sec:Conclusions} we conclude and summarise our findings.

\section{Observations}\label{sec:Observations}
\begin{figure*}
    \centering
        \includegraphics{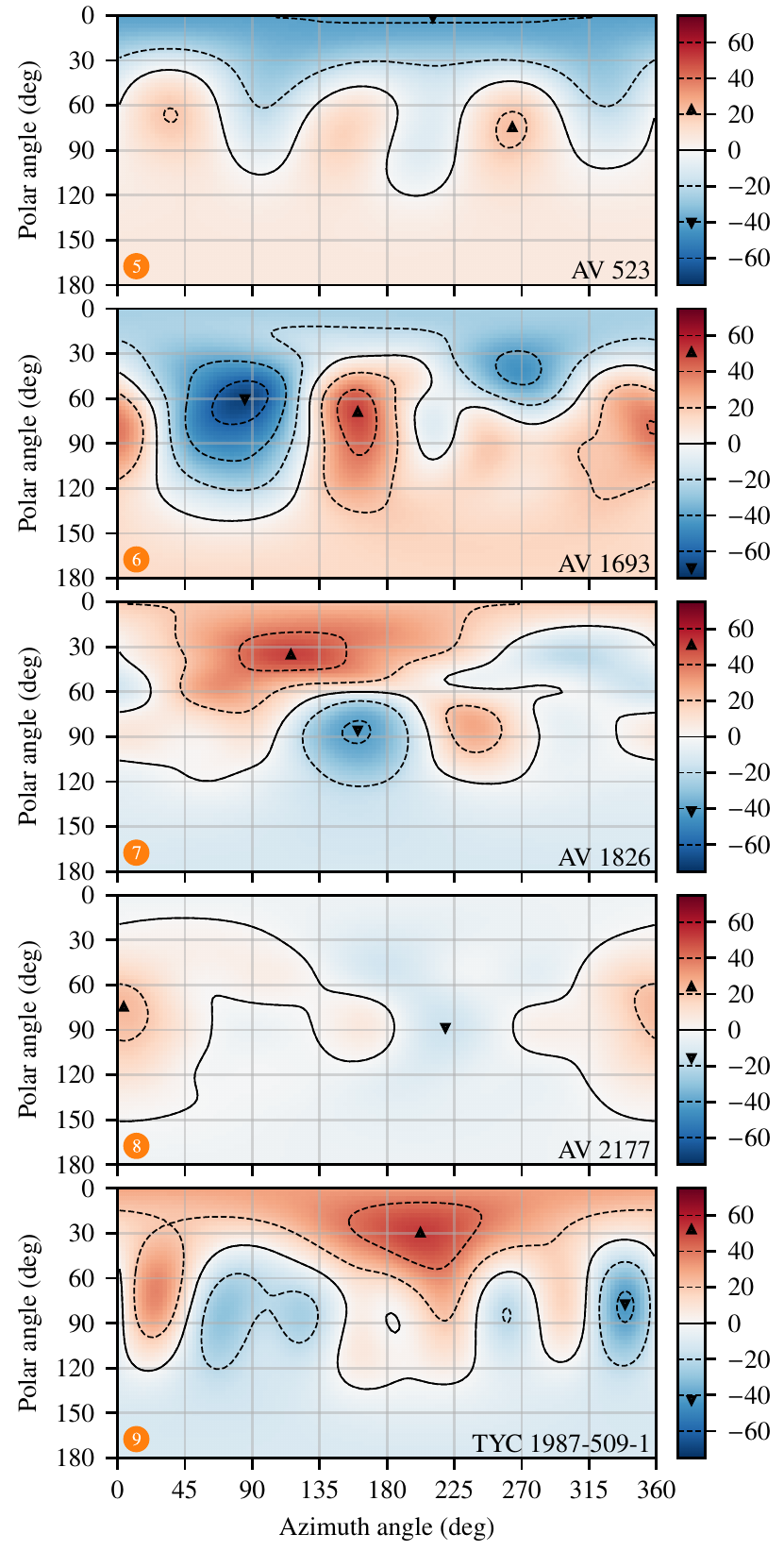}
        \includegraphics{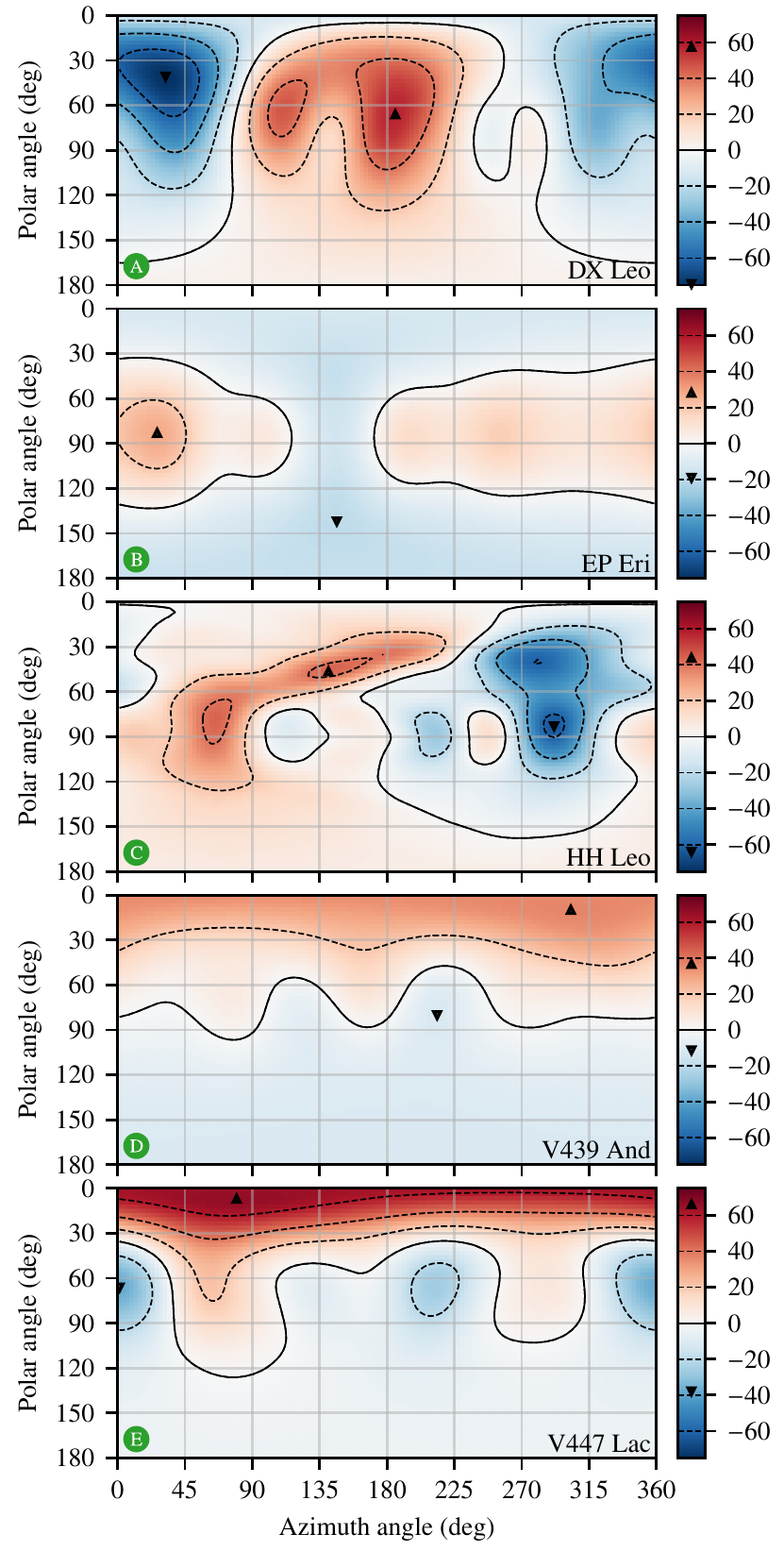}
    \caption{
        Radial magnetic field strength in Gauss for the stars modelled in this work based on the radial magnetic field coefficients derived in~\citetalias{2016MNRAS.457..580F} and \citetalias{2018MNRAS.474.4956F}. 
        The polar angle is measured from the rotational north pole, while the azimuth angle is measured around the stellar equator. The smallest scale of representable features is \SI{12}{\degree} as described in Section~\ref{sec:mag-map-zdi}.        
        The fully drawn contour line indicates a zero value of radial magnetic field strength, and the dashed contour lines represent increments as shown in the corresponding colour bar on the right of each plot. 
        The upwards-pointing and downwards-pointing black triangles show the position and value of the maximum and minimum radial field strength values; they also appear in the colour bars so that the range of values may be easily discerned.
    }\label{fig:magnetograms-zdi}
\end{figure*}

In this work we use stellar surface magnetic field maps based on spectropolarimetric observations of the Hercules-Lyra association made with the 
Narval instrument~\citep{2003EAS.....9..105A} at the Télescope Bernard Lyot, 
and of the Coma Berenices cluster made with the 
ESPaDOnS instrument~\citep{2003ASPC..307...41D,2012MNRAS.426.1003S} at the Canada-Hawaii-France Telescope. Both sets of observations were part of the `TOwards Understanding the sPIn Evolution of Stars' (\toupies{}) project\footnote{\url{http://ipag.osug.fr/Anr_Toupies/}}; the ESPaDOnS observations were also part of the \emph{History of the Magnetic Sun} Large Program at the CFHT.\@ The reduced spectra associated with \toupies{} are available from the Polarbase \citep{1997MNRAS.291..658D,2014PASP..126..469P} website\footnote{\url{http://polarbase.irap.omp.eu/}}.

During the observations the instruments recorded the Stokes \(V\) and Stokes \(I\) circular polarisation spectrum and total intensity spectrum over a period of a few weeks. The time period minimises actual variations in the stellar magnetic field and provides sufficient phase coverage to map the entire visible surface. The stellar magnetograms that were produced from these observations are published in~\citetalias{2016MNRAS.457..580F}~and~\citetalias{2018MNRAS.474.4956F}; we refer the readers to these papers for a detailed description of the observations.  Table~\ref{tab:observed_quantities} give the stars' fundamental parameters, and Fig.~\ref{fig:magnetograms-zdi} shows the radial magnetic field components of the magnetograms.

In Section~\ref{sec:Discussion} we also include our previous Solar and Hyades cluster wind models from \citetalias{paper1}; the magnetograms used to drive those models were published in 
\citetalias{2018MNRAS.474.4956F}.

\subsection{Magnetic mapping with Zeeman-Doppler imaging}\label{sec:mag-map-zdi}
Zeeman-Doppler imaging \citep[ZDI,][]{1989A&A...225..456S} has been used successfully to image the large-scale surface magnetic field of cool stars, see the review by~\citet{2009ARA&A..47..333D}. By using least square deconvolution~\citep[LSD,][]{1997MNRAS.291..658D,2010A&A...524A...5K} individual circularly polarised spectral lines are combined into a single LSD profile with a high signal-to-noise ratio. 

In modern ZDI the observable surface magnetic field is decomposed into a set of spherical harmonic coefficients~\citep{1999MNRAS.305L..35J,2006MNRAS.370..629D}. The coefficients are found by applying the maximum entropy image reconstruction~\citep{1984MNRAS.211..111S} method to the LSD profile. In this way a set of coefficients is found that satisfies a \(\chi^2\) bound on the fit while maximising an entropy measure; we direct the reader to~\citetalias{2016MNRAS.457..580F} and~\citetalias{2018MNRAS.474.4956F} for a detailed description of this process, which recovers all three vector components of the surface magnetic field. This work only makes use of the radial field components as is common in MHD wind modelling (see Section~\ref{sec:numerical_model}). The stellar surface radial magnetic field is represented as the real part of an orthogonal sum of the form
\begin{equation}\label{eq:zdi_br}
    B_r(\theta, \varphi) =
    \sum_{\ell=1}^{\ell_\text{max}} \,
    \sum_{m=0}^\ell \alpha_{\ell m} 
    \sqrt{\frac{2\ell+1}{4\pi} \frac{(\ell-m)!}{(\ell+m)!}} 
    P_{\ell m}(\cos \theta) e^{i m\varphi} 
\end{equation}
where \(\alpha_{\ell m} \) are the complex-valued spherical harmonics coefficients of the radial field, \(P_{\ell m}(\cos \theta)\) is the associated Legendre polynomial of order \(m\) and degree \(\ell\),
and \(\theta, \varphi\) are the polar and azimuthal angles which identify points on the stellar surface. Since only the real part of equation~\eqref{eq:zdi_br} is of interest negative \(m\) values are omitted from the sum\footnote{Negative values of \(m\) would provide redundant degrees of freedom as the imaginary part equation \eqref{eq:zdi_br} is discarded.}.   

The smallest features that can be reproduced by equation~\eqref{eq:zdi_br} are \( \mysim \SI{180}{\degree}/\ell_\text{max}\) in angular diameter; the magnetograms in this work use \(\ell_\text{max}=15\) so that the smallest representable feature scale is \SI{\mysim 12}{\degree}.
The `effective' degree \(\ell_\text{eff}\) can however be significantly lower than \(\ell_\text{max}\), resulting in only large scale features being present in the ZDI magnetograms. 
The effective degree depends on observational parameters such as the unpolarised line width, the star's projected rotational velocity \(v \sin i\), and the observations' signal-to-noise ratio  \citep{1997A&A...326.1135D,2010MNRAS.407.2269M}.
We estimate the effective degree, \(\ell_{.90}\) and \(\ell_{.99}\), as the value of \(\ell\) where \SI{90}{\percent} and \SI{99}{\percent} of the magnetic field energy is stored in degrees lower than or equal to \(\ell\); the resulting values are given in Table~\ref{tab:magnetic_averages}, where we see values from 2 to 5 for \(\ell_{.90}\) and values from 4 to 8 for \(\ell_{.99}\). 
Note that the surface magnetic field values in Table~\ref{tab:magnetic_averages} are calculated from the steady state magnetic field in our model output and as such they do not directly correspond to the magnetic averages in \citetalias{2016MNRAS.457..580F,2018MNRAS.474.4956F}. This is because the perpendicular components of the surface magnetic field are free to settle and converge with the numerical solution  (see Section~\ref{sec:numerical_model}) instead of being held at their ZDI-derived values, which partly originate from photospheric currents. 
The \(\ell_{.90}\) and \(\ell_{.99}\) values do still give a good indication of the magnetic field complexity in a single parameter. 
\begin{table*}
    \centering
    \caption{
        Aggregate surface magnetic field values. \(|B_r|\) and \(\max |B_r|\) are the surface average and maximum absolute radial field values in Fig.~\ref{fig:magnetograms-zdi}, 
        i.e.\ the average value of \(|B_r(\theta, \varphi)|\) over the stellar surface, and the maximum value of \(|B_r(\theta, \varphi)|\) over the stellar surface.  The location \((\theta_\text{max}, \varphi_\text{max})\) of \(\max|B_r|\) on the stellar surface can be seen in Fig.~\ref{fig:magnetograms-zdi}.
        \(|\vec B|\) is the average surface field strength of the final steady-state solution,
        i.e.\ the average of \(|\vec B(\theta, \varphi)|\) over the stellar surface.
        `Dip.', `Quad.', `Oct.' and `16+' are the final fraction of magnetic energy in dipolar (\(\ell=1\)), quadrupolar (\(\ell=2\)), octupolar (\(\ell=3\)) and hexadecapolar and higher (\(\ell\geq 4\)) modes. The \(\ell_{.90}\) and \(\ell_{.99}\) columns refer to the magnetogram degree at which \SI{90}{\percent} and \SI{99}{\percent} of the magnetogram energy is contained in degrees lower than or equal to the tabulated value.
        Note that the tabulated \(|\vec B|\) values and percentages do not correspondence directly with the photospheric values in \citetalias{2016MNRAS.457..580F} and \citetalias{2018MNRAS.474.4956F}.
    }\label{tab:magnetic_averages}
    \sisetup{
        table-figures-decimal=1,
        table-figures-integer=3,
        table-number-alignment=center,
        round-mode=places,
        round-precision=1}
    \input{tables/aggregate-magnetic-quantities.tex}
\end{table*}

While the ZDI method does not provide uncertainty estimates along with the magnetic maps, it is accepted that ZDI is able to reproduce the structure of the large-scale magnetic field.
This is supported by the study of \citet{2000MNRAS.318..961H}, which found similar results using different ZDI implementations.
Polarity reversals have been observed for the stars
\(\tau\)~Bo\"{o}tis~\citep{2008MNRAS.385.1179D,2009MNRAS.398.1383F,2013MNRAS.435.1451F,2016MNRAS.459.4325M} and HD~75332~\citep{2021MNRAS.501.3981B}. Evidence of field reversals in HD~78366 and HD~190771 was found by \citet{2011AN....332..866M}; the authors also found evidence of a more complex cycle in \(\xi\)~Bo\"{o}tis~A.
Different ZDI implementations do not always agree on details of the medium- and small-scale magnetic field as noted in the review of~\citet{2016LNP...914..177K}, however it is the large scale field that shapes the coronal magnetic field.

\section{Simulations}\label{sec:Simulations}
In this section we provide an overview of the numerical simulations carried out as part of this work.

\subsection{Model equations}\label{sec:model-equations}
We use the 
Alfvén Wave Solar Model~\cite[\awsom{}, ][]{2013ApJ...764...23S, 2014ApJ...782...81V} of the
Space Weather Modelling Framework~\cite[\swmf{}, ][]{2005JGRA..11012226T,2012JCoPh.231..870T}
to produce wind models driven by the radial component of the \toupies{} magnetic maps as described in Section~\ref{sec:mag-map-zdi}. The \awsom{} model is built upon the \batsrus{} model \citep{1999JCoPh.154..284P,2012JCoPh.231..870T}. An overview of \awsom{} is found in the review of~\citet{2018LRSP...15....4G}.

Alfvén waves are a mechanism of coronal heating~\citep{1968ApJ...154..751B} that has been thought to contain enough energy~\citep{1968ApJ...153..371C} to power the Solar wind. 
In the \awsom{} model the wind is heated to coronal temperatures by Alfvén wave energy originating in deeper stellar layers; this is modelled as a Poynting flux \(\Pi_\Alfven\) proportional to the local \(|\vec{B}|\) value at the inner model boundary. The two-temperature MHD equations are thus extended to model the propagation and dissipation of Alfvén wave energy.

\begin{subequations}
We briefly state the set of differential equations solved by \awsom{}; for a detailed description including the cooling and heating terms we refer the reader to \citetalias{paper1}. Mass conservation is given by 
\begin{equation}
    \label{eq:mass_conservation}
    \frac{\partial \rho}{\partial t} + \nabla \left(\rho \vec{u}\right) = 0, 
\end{equation}
where \(\rho\) is the mass density and \(\vec{u}\) is the flow velocity.
The induction equation is
\begin{equation}
    \label{eq:induction}
    \frac{\partial \vec{B}}{\partial t} + \nabla \left(\vec{u}\vec{B} - \vec{B}\vec{u}\right) = 0,
\end{equation}
where \(\vec{B}\) is the magnetic field.
The time evolution of Alfvén wave energy density in parallel (\(w^+\)) and antiparallel (\(w^-\)) directions along the magnetic field is described by
\begin{equation}
    \label{eq:alfven_waves}
    \frac{\partial w^\pm}{\partial t} + \nabla \left((\vec{u} \pm \vec{v}_\Alfven ) w^\pm \right) + \frac{w^\pm}{2}\left(\nabla \cdot \vec{u}\right) 
    = \mp R \sqrt{w^{-} w^+} - Q^\pm_{\Wind}
\end{equation}
where \(\vec{v}_\Alfven= \vec{B}/\sqrt{\mu_0 \rho}\) is the Alfvén velocity, \(\mp R \sqrt{w^{-} w^+}\) are reflection rates that transfer energy between \(w^+\) and \(w^-\), 
and \(Q^\pm_{\Wind}\) is a dissipation term. 
Momentum conservation is given by
\begin{equation}
    \label{eq:momentum_conservation}
    \frac{\partial \left(\rho \vec{u}\right)}{\partial t} 
    + \nabla \left( \rho \vec{u}\vec{u} - \frac{\vec{B}\vec{B}}{\mu_0}
    + P
    + \frac{B^2}{2\mu_0} + P_\Alfven \right)
    = -\rho \frac{GM \vec{r}}{r^3}
\end{equation}
where \(P=P_\text{i} + P_\text{e}\) is the sum of the ion and electron thermal pressure and \(P_\Alfven = \left.\left(w^+ + w^-\right)\right/2\)  is the Alfvén wave pressure. The constants \(G\) and \(\mu_0\) are the gravitational constant and the vacuum permeability, \(M\) is the stellar mass, and \(\vec{r}\) is the positional vector (relative to the stellar centre).
The ion energy equation is 
\begin{equation}
    \label{eq:pressure_ion}
    \frac{\dfrac{\partial P_\text{i}}{\partial t} + \nabla \left(P_\text{i}\vec{u}\right) }
    {\left(\gamma -1\right) }
    + P_\text{i}\nabla\vec{u} =
        \frac{P_\text{e}-P_\text{i}}{\tau_\text{eq}} 
    + f_\text{i} Q_\Wind
    - \rho \frac{GM \vec{r} \cdot \vec{u} }{r^3},
\end{equation}
and the electron pressure equation is
\begin{equation}
    \label{eq:pressure_electron}
    \frac{\dfrac{\partial P_\text{e}}{\partial t} + \nabla \left(P_\text{e}\vec{u}\right)}{\gamma-1} 
    + P_\text{e}\nabla\vec{u}
    = 
    \frac{P_\text{i}-P_\text{e}}{\tau_\text{eq}} 
    + f_\text{e}Q_\Wind
    -Q_\text{rad}
    -\nabla\vec{q}_\text{e},
\end{equation}
with \(\gamma=5/3\) being the ratio of specific heats for monatomic gases. The right hand side terms are 
collisional energy transfer \(\pm (P_\text{i}-P_\text{e})/\tau_\text{eq}\);
ion/electron heating by Alfvén wave dissipation \(f_\text{i}Q_\Wind\) and\(f_\text{e}Q_\Wind\) where \(f_\text{i}+f_\text{e}=1\);
work against gravity \(\rho G M \vec r \cdot \vec u / r^3\); 
radiative losses \(Q_\text{rad}\); and 
electron heat conduction \(\nabla \vec q_\text{e}\). 
\citetalias{paper1} gives detailed descriptions of these terms.
\end{subequations} 

\subsection{Numerical model and boundary conditions}\label{sec:numerical_model}
The model domain comprises two partially overlapping regions. The inner region uses a spherical grid and the outer region uses a Cartesian grid. The mesh is selectively refined near the stellar surface and the current sheet region where the sign of \(B_r\) changes. By stepping forward in time a steady state is reached where the magnetic and hydrodynamic forces are in balance. 

As was done in \citetalias{paper1} we attempt to control for the uncertainty associated with the surface magnetic field strength measured by ZDI, we conduct two series of simulations denoted \(B_\ZDI\) and \(5B_\ZDI\). 
The two series of models are identical, except that the surface radial magnetic field strength is increased by a factor of 5 in the \(5B_\ZDI\) series. 
As the octree-based grid refinement occurs near the current sheet, the refined grid may be slightly different between the \(B_\ZDI\) and \(5B_\ZDI\) case when the position of the current sheet itself is differing between the cases; we do not expect this to have any influence on the model results. 
The case names of each individual model in the \(B_\ZDI\) series is given in Table~\ref{tab:observed_quantities}; these names are used throughout this paper to identify the individual models. 
In the \(B_\ZDI\) series the models are denoted as e.g. `\mbox{\Unscaled{8}{C1}AV~2177}'; a numbered circle and a short form of the star name. The corresponding model in the \(5B_\ZDI\) is denoted as `\mbox{\Scaled{8}{C1}$5\times$AV~2177}'; a numbered star and $5\times$ followed by the star's case name.

We use the same model parameters as in \citetalias{paper1}: 
The temperature and number density at the chromospheric inner boundary is set to Solar values \(T=\SI{5e4}{\kelvin}\) and \(n=\SI{2e17}{\per\cubic\meter}\) 
similarly to~\citet{2016A&A...588A..28A} for example. 
The outgoing Alfvén wave energy density at the inner boundary is set to 
\(w = (\Pi_\Alfven/B)_\Sun\sqrt{\mu_0 \rho} \)~\citep{2014ApJ...782...81V}. %
The value \((\Pi_\Alfven/B)_\Sun=\SI{1.1e6}{\watt\per\square\meter\per\tesla}\)~\citep{2018LRSP...15....4G} is calibrated so that \awsom{} reproduces Solar wind conditions.
A corrective scaling \((\Pi_\Alfven/B) = (\Pi_\Alfven/B)_\Sun (R/R_\Sun)^{0.3}\) as suggested by \citet{2013ApJ...764...23S}
has previously been employed by \citet{2016ApJ...833L...4G,2018PNAS..115..260D} in M-dwarf wind modelling, see also \citet{2021LRSP...18....3V}.
That corrective scaling is not applied as it would change the value of \((\Pi_\Alfven/B)\) by less than \SI{10}{\percent} for the stellar radii in this study; 
we find that the two smallest \(R\) stars \Unscaled{5}{C1}AV~523 and \Unscaled{A}{C2}DX~Leo give wind mass loss values of \SI{\mysim 80}{\percent} of their unscaled values and similar or lower variation for the other parameters of interest. 

Recently it has been shown that wind mass loss \(\dot M\) is roughly proportional to \(\Pi_\Alfven/B\) \citep{2020A&A...635A.178B} for Solar wind models, and some authors such as \citet{2021ApJ...916...96A} have applied large scalings to the parameter when modelling young, Solar-type stars. We briefly return to this issue and its implications in Section~\ref{sec:Conclusions}.

The radial component of the boundary magnetic field is fixed to the local magnetogram value in Fig.~\ref{fig:magnetograms-zdi}, i.e.\ \(\vec{B}_\ZDI\cdot \uvec{r}\) or \(5\vec{B}_\ZDI\cdot \uvec{r}\), depending on the model series. The non-radial surface magnetic field \(\vec{B}_\perp\) is part of the steady state solution inside the model domain.

\section{Results}\label{sec:results}
In this section we give an overview of the features in each wind model, focusing on 
the coronal magnetic field structure in Section~\ref{sec:coronal_structure};
the Alfvén surface and wind speed in Section~\ref{sec:alfven_surface};
the wind mass loss and angular momentum loss in Section~\ref{sec:mass-loss-angmom-loss};
and the wind pressure in Section~\ref{sec:wind-pressure}.
Having the full three-dimensional wind solutions make it possible to calculate a large range of wind-related quantities, 
including wind mass loss, angular momentum loss, and wind pressure for an Earth-like planet. These parameters and others are presented in Table~\ref{tab:main-table}.  

\subsection{Coronal structure}\label{sec:coronal_structure}
\begin{figure*}
    \centering
    \includegraphics{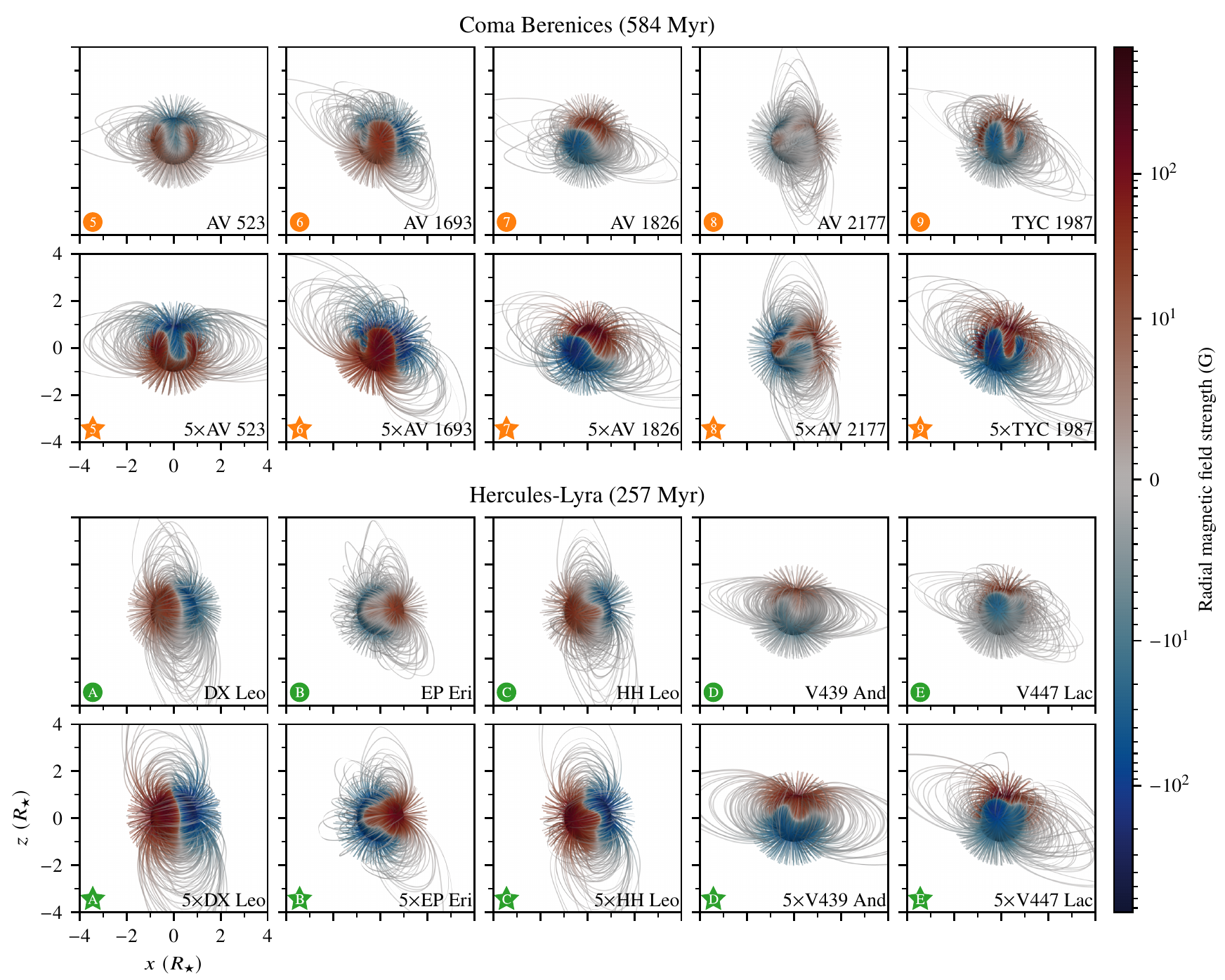}
    \caption{
        This plot shows the magnetic field structure in the stellar coronae for the final, relaxed wind solutions. The top two rows show the unscaled \(B_\ZDI\) (upper) and scaled \(5B_\ZDI\) (upper middle) series wind models for the Coma Berenices wind models. The bottom two rows show the unscaled (lower middle) and scaled (lower) models for the Hercules-Lyra models. A selection of magnetic field lines are shown; the open magnetic field lines are truncated at \(4R_\Star\) to avoid crowding out the region of closed magnetic field. The stellar surface and the magnetic field lines are coloured by the local value of the radial magnetic field. The colour scale is linear from \SIrange{-10}{10}{\gauss} and logarithmic outside of this range. In each plot the stellar axis of rotation coincides with the plot \(\uvec z\) axis, and the rotational phase shown is chosen in order to permit the easy discrimination of the regions of open and closed magnetic field lines. 
        The structure of the coronal magnetic fields appears dipole-like in spite of the excursions from a dipolar structure of the surface radial magnetic field.
    }\label{fig:paper2-br-fieldlines}
\end{figure*}

The structure of the coronal magnetic field is shown in Fig.~\ref{fig:paper2-br-fieldlines}. Open field lines are not indicated past four stellar radii. In each panel the stellar surface and the field lines are coloured by their radial magnetic field strength. The colour scale used is linear in the range 
\SIrange[]{-10}{10}{\gauss} 
and logarithmic outside this range as indicated by the position of the tick marks on the figure colour scale. 

From Fig.~\ref{fig:paper2-br-fieldlines} it is clear that the coronal field tends towards a dipole-like structure in spite of the differences in surface magnetic field geometry, and large excursions from a dipolar structure by the radial surface magnetic field \(B_r\) that may be seen in Fig.~\ref{fig:magnetograms-zdi} and Fig.~\ref{fig:paper2-br-fieldlines}. This tendency indicates that the dipolar magnetogram coefficients \(\alpha_{10}\) and \(\alpha_{11}\), for which the degree \(\ell=1\) in equation~\eqref{eq:zdi_br}, largely govern the shape of the coronal field as one moves away from the stellar surface. This is in agreement with the complementary method of potential extrapolations of the surface magnetic field into the corona
\citep{1969SoPh....6..442S,1969SoPh....9..131A,1984PhDT.........5H,1992ApJ...392..310W} where higher degree terms decay more rapidly with increasing \(|\vec r|\).

At the same time it should be kept in mind that, in contrast to the tidy dipolar coronal fields of Fig.~\ref{fig:paper2-br-fieldlines}, the Sun's magnetic large-scale field does not always resemble a dipole, especially around periods of high activity. This hints to the importance of the missing medium- and small-scale magnetic field in state-of-the-art stellar magnetograms in reconstructing the stellar coronal structure. 

To emphasise the dipole-like nature of the coronal field in our wind models the rotational phase of each stellar model is chosen so that the plotted projected angle between the magnetic dipole axis and \(\uvec z\) is maximised and the dipole appears side-on in each panel of the figure, accentuating the structure of open and closed magnetic field lines. 
As in \citep{2009ApJ...699..441V} and \citetalias{paper1} we see that shape of the closed field region terminates in so-called `helmet' streamers named after spiked military helmets \citep[see e.\ g.][]{knotel1980uniforms}.

While we note a general agreement between the coronal field found via potential field extrapolation methods and the relaxed fields found in our MHD models, the steady state surface magnetic field of the MHD solution differs from the input ZDI field as given in \citetalias{2016MNRAS.457..580F} and \citetalias{2018MNRAS.474.4956F} as the ZDI field includes magnetic field originating from photospheric currents. In our models the non-radial surface magnetic field 
\(
    \vec B_\perp = B_\theta \uvec{\theta} + B_\varphi \uvec{\varphi}
\) 
is found from the final, relaxed state of the MHD solution; the absence of photospheric currents in our model means that the resulting fields have only small non-potential components at the stellar surface. \citet{2013MNRAS.431..528J} observed in their models that the non-potential field has little influence on the large-scale magnetic geometry in the corona, and as such it is also not expected to influence the steady state stellar wind. The non-potential field does, however, represent a source of available energy to power transient expulsions of plasma and magnetic energy. %

In Table~\ref{tab:magnetic_averages} we give some aggregate quantities of the surface magnetic field.
The \(|B_r|\) column represents the average radial field strength at the stellar surface:
\begin{equation}
    |B_r| = \frac{1}{4\pi R_\Star^2}\oint\nolimits_{S_\Star} \left|\vec B \cdot \uvec n\right|\,\mathrm{d}S= \frac{1}{4\pi R_\Star^2}\oint\nolimits_{S_\Star} \left|B_r(\theta, \varphi)\right|\,\mathrm{d}S
\end{equation}
where \(S_\Star\) is the stellar surface and \(\uvec n \parallel \uvec r\) is the normal vector of \(S_\Star\). 
The \(\max |B_r|\) column gives the maximum value of the radial field strength over the stellar surface:
\begin{equation}
    \max |B_r| = \max_{S_\Star} \left|\vec B\cdot \uvec n\right| =  \max_{\theta, \varphi} \left|B_r(\theta, \varphi)\right|
\end{equation}
As the radial surface magnetic field strength is a fixed boundary condition of our model, the values of \(|B_r|\) and \(\max |B_r|\) do not change as the wind model is relaxed towards a steady state.
    The \(|\vec B|\) column gives the average surface field strength; the value is calculated as
    \begin{equation}
        |\vec B| 
        = \frac{1}{4\pi R_\Star^2}\oint\nolimits_{S_\Star}  
        \left|\vec B(\theta, \varphi)\right| 
        \, \mathrm{d}S;
    \end{equation}
    the value of \(|\vec B|\) can vary during the relaxation process as \(\vec B = B_r\uvec r + \vec B_\perp\), of which the latter term is not a fixed boundary condition of the model. As expected, we observe that the relaxed value of \(\vec B\) is similar to the `poloidal' magnetic field strength in \citetalias{2016MNRAS.457..580F,2018MNRAS.474.4956F}, which is unaffected by photospheric currents.
The `Dip', `Quad', `Oct' `16+' columns give the percentage of magnetic energy in dipolar, quadrupolar, octupolar, and hexadecapolar and higher models in the relaxed surface magnetic field. 
The \(\ell_{.90}\) and \(\ell_{.99}\) columns give the degree for which \SI{90}{\percent} and \SI{99}{\percent} of the magnetic energy is contained in degrees equal to or lower than the tabulated value. 

For each stellar model we also compute numerically the total absolute magnetic flux at the stellar surface, as well as measures of the so-called open and axisymmetric magnetic fluxes. 
The absolute magnetic flux across a closed surface \(S\) is given by 
\begin{equation}
\Phi(S) = \oint\nolimits_S |\vec B \cdot \uvec n| \, \mathrm{d}S.
\end{equation}
where \(\uvec n\) is the normal vector of \(S\).
We calculate the unsigned magnetic flux at the stellar surface, which we denote \(\Phi\), and for which we also have the analytical result \(\Phi = 4\pi R^2 |B_r|\). The values of \(\Phi\) are provided in Table~\ref{tab:main-table}. 
As \(\Phi\) is calculated from the fixed radial magnetic field only, it does not change with time in our solution.

The total absolute magnetic flux at the stellar surface may be thought of as comprising a closed flux \(\Phi_\text{closed}\) of `closed' magnetic field lines that have two foot-points on the stellar surface, and an `open' flux \(\Phi_\text{open}\) of magnetic field lines that have only one such foot-point. In contrast to a magnetic multipole in vacuum, these open field lines extend into space, dragged by the escaping stellar wind. Closed magnetic field lines are plentiful near the stellar surface, while \(\Phi_\text{open} \gg \Phi_\text{closed}\) at large distances from the star. We calculate \(\Phi_\text{open}\) for each model by numerically integrating \(\Phi(S)\) for a spherical surface \(S\) with a radius of many Solar radii. The resulting measure of the open flux is not sensitive to the exact radius used. The resulting values of \(\Phi_\text{open}\) are given in Table~\ref{tab:main-table}.

The axisymmetric open flux has been linked to the cosmic ray intensity at the Earth's surface \citep{2006ApJ...644..638W} through the formation of regions of higher magnetic field strength. 
The axisymmetric open flux is the integral over a spherical surface of the azimuth-averaged value of \(\vec B\) \citep{2014MNRAS.438.1162V}: 
\begin{equation}
    \label{eq:openflux}
    \Phi_\text{axi} = \oint\nolimits_S |\vec B_\text{axi} \cdot \uvec n| \,\mathrm{d}S, 
    \text{ where }
    \vec B_\text{axi} = \frac{1}{2\pi}\oint \vec B \,\mathrm{d}\varphi. 
\end{equation}
The numerical value of \(\Phi_\text{axi}\) is found by integrating equation \eqref{eq:openflux} over the same surface \(S\) as is used when finding \(\Phi_\text{open}\). 
A quantity related to \(\Phi_\text{open}\) is the open surface fraction \(S_\text{open}/S\), the fraction of the stellar surface crossed by open magnetic field lines.
We find this quantity by tracing the magnetic field lines from large number of evenly sampled points on the stellar surface. 

The values of \(S_\text{open}\) in Table~\ref{tab:main-table} confirms our impression that the region of closed field lines is greater for the \(5B_\ZDI\) series of models both for the Coma Berenices cluster and the Hercules-Lyra association. 
The region of open magnetic field lines appears about \SI{\mysim35}{\percent} smaller for the models in the \(5B_\ZDI\) series than for the models in the \(B_\ZDI\) series. This affects the wind density and speed as the regions of fast stellar wind (open field lines) are reduced; we investigate the effect of surface magnetic field strength on wind mass loss, etc.\ in Section~\ref{sec:effect-of-magnetic-scaling}.

\subsection{Alfvén surface and current sheet}\label{sec:alfven_surface}
\begin{figure*}
    \centering
    \includegraphics{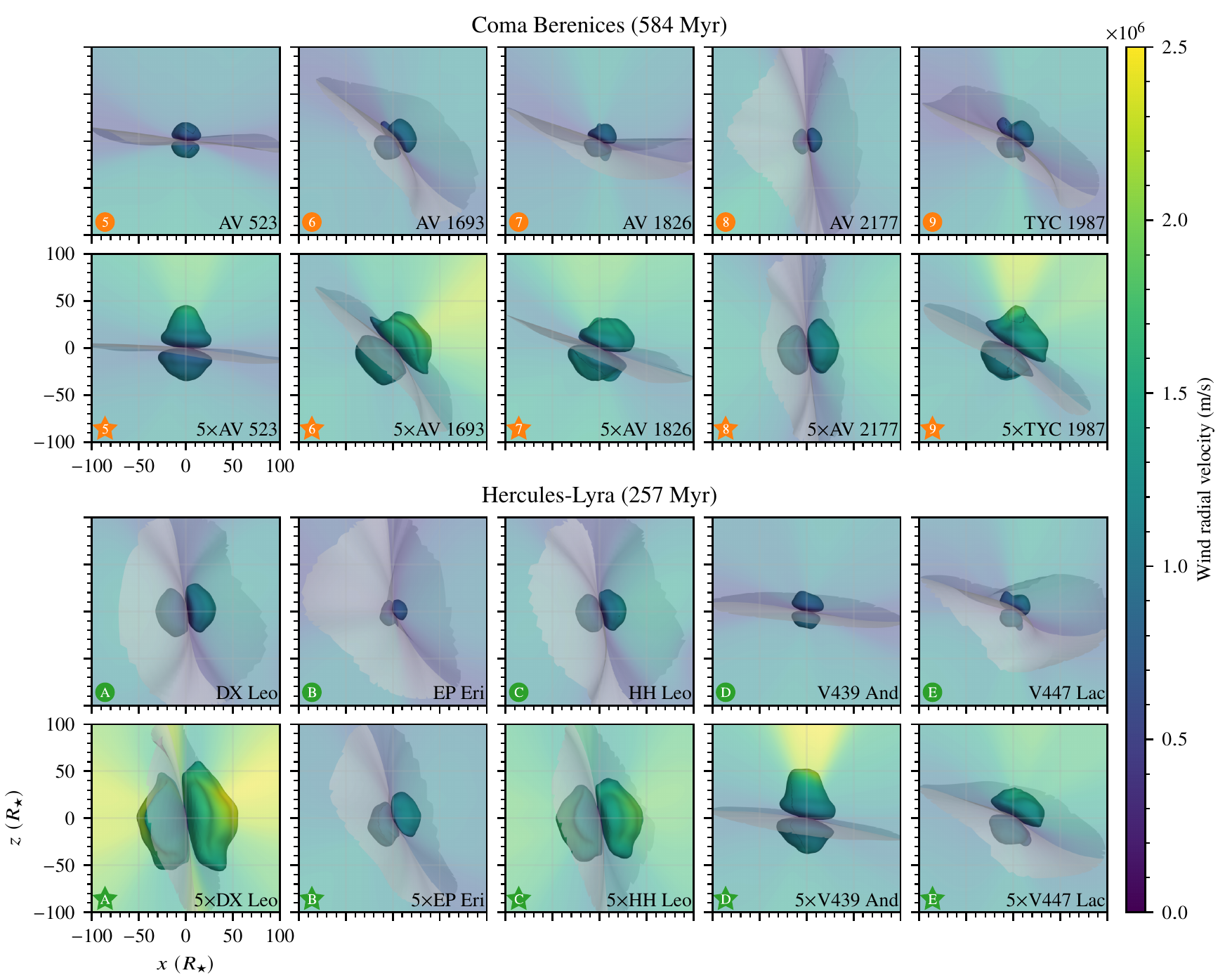}
    \caption{
        Alfvén surface and current sheet for the Coma Berenices stars (top) and the Hercules-Lyra stars bottom. The order of the panels, which is the same as in Fig.~\ref{fig:paper2-br-fieldlines} is indicated by the bottom left symbols in each panel. In each group the unscaled \(B_\ZDI\) models are shown in the top row and the scaled \(5B_\ZDI\) are shown in the bottom row. 
        The \(z\) axis is parallel to the stellar axis of rotation \(\uvec{z} \parallel \vec{\Omega}\). 
        The current sheet, for which \(B_r=0\) is shown as a translucent grey surface, truncated at \(100 R_\Star\).
        The plane of sky (\(xz\) plane) and the Alfvén surface are coloured according to the local wind radial velocity.
        In general, we observe that the amplified magnetic fields of the \(5B_\ZDI\) models produce more irregular Alfvén surfaces.  
    }\label{fig:paper2-ur-alfven}
\end{figure*}
\begin{table*}
    \centering
    \caption{
        Aggregate wind quantities calculated from the wind models in Figs.~\ref{fig:paper2-br-fieldlines}~to~\ref{fig:paper2-Pwind-IH-equatorial}. 
        The unsigned surface magnetic flux \(\Phi = 4\pi R^2 |B_r|\) represents the absolute amount of magnetic flux crossing the stellar surface. 
        The open magnetic flux \(\Phi_\text{open}\) represents the amount of surface flux contained in regions of open magnetic field lines. 
        The axisymmetric flux \(\Phi_\text{axi}\) is the axisymmetric part of the open flux.
        The open surface fraction \(S_\text{open} / S\) is the fraction of the stellar surface crossed by open magnetic field lines.
        The average Alfvén radius \(R_\Alfven\) is the radial distance to the Alfvén surface averaged over the stellar surface.
        The torque-averaged Alfvén radius \(|\vec r_\Alfven\times \uvec \Omega|\) is the torque arm length at the Alfvén surface, averaged over the stellar surface.
        \(\dot M\) and \(\dot J\) are the mass and angular momentum carried away by the wind. 
        \(P_\Wind^\Earth\) is the average wind pressure for an Earth-like planet, averaged over solar and orbital phase, and 
        the magnetospheric stand-off distance for an Earth-like planet 
        \(R_\text{m}/R_\text{p}\) is the corresponding distance, in planetary radii, from the centre of the Earth-like planet to the region where the stellar wind encounter the magnetosphere of the Earth-like planet. 
    }\label{tab:main-table}
    \sisetup{
        table-figures-decimal=1,
        table-figures-integer=1,
        table-figures-uncertainty=0,
        table-figures-exponent = 3,
        table-number-alignment=center,
        round-mode=places,
        round-precision=1
    }
    \input{tables/main-table-body.tex}

\end{table*}
When the wind speed exceeds the Alfvén speed \(u_\Alfven = B / \sqrt{\mu_0 \rho}\) information cannot propagate against the flow direction. 
The Alfvén surface \(S_\Alfven\) comprises the points where the local wind velocity \(u = u_\Alfven\). 
As the wind accelerates with increasing distance from the star in a non-uniform way determined by the surface magnetic field, the exact shape of the Alfvén surface is determined by the magnetogram 
and the parameters of the coronal heating model (as described in Section~\ref{sec:numerical_model}).
Fig.~\ref{fig:paper2-ur-alfven} shows the resulting Alfvén surface for each of the coronal fields in Fig.~\ref{fig:paper2-br-fieldlines}. 
The orientation of the plots are the same as in Fig.~\ref{fig:paper2-br-fieldlines}. 
The current sheet \citep{1971CosEl...2..232S} where \(B_r=0\) is shown as a grey surface in Fig.~\ref{fig:paper2-ur-alfven}. 
The inclination of the inner current sheet with respect to the stellar axis of rotation \(i_{B_r=0}\) (reported in Table~\ref{tab:main-table}) is found by fitting a plane to the inner current sheet and computing the angle between the fitted plane normal vector and the \(\uvec \Omega\) axis.

From the plots of the Alfvén surfaces and inner current sheets in Fig.~\ref{fig:paper2-ur-alfven} we observe two-lobed Alfvén surfaces with a variety of inclinations with respect to the stellar axis of rotation. 
We do not observe any qualitative differences between the models of the Coma Berenices stars and the models of the Hercules-Lyra stars. 
Quantitative differences are considered in Section~\ref{sec:Discussion}. 
As was the case in \citetalias{paper1} we do observe clear qualitative differences between the unscaled models and the scaled models in both Coma Berenices and Hercules-Lyra; 
the scaled surface magnetic field gives rise to larger Alfvén surface lobes and consequently greater values of \(R_\Alfven\) and  
\(|\vec r_\Alfven \times \uvec \Omega|\). In addition to being larger, the Alfvén surface lobes of the scaled models also tend to be more irregular, and give rise to greater wind velocities. 

The large range of `magnetic inclinations' in both the \(B_\ZDI\) and \(5B_\ZDI\) is evident both in Figure~\ref{fig:paper2-ur-alfven} and in the \(i_{B_r=0}\) values of Table~\ref{tab:main-table}. The stars AV~523 and V439~And have current sheets that are are nearly aligned to the axis of rotation \(\uvec \Omega\), with \(i_{B_r=0}\sim\SI{4}{\degree}\) and \(i_{B_r=0}\sim\SI{10}{\degree}\) respectively, while AV~2177, DX~Leo, EP~Eri and HH~Leo have  \(i_{B_r=0}\gtrsim\SI{75}{\degree}\). The variation in \(i_{B_r=0}\) gives rise to a corresponding range of inclination of Alfvén surface lobes as can be seen in Fig.~\ref{fig:paper2-ur-alfven}. 

Looking at individual wind models, we observe that the two scaled models 
\Scaled{5}{C1}AV~523
and
\Scaled{D}{C2}V439~And
exhibit large Alfvén surface radii near the rotational north pole (\(+z\) direction) giving the northern Alfvén lobes of these two stars a noticeable ovoid shape also seen in \citet{2014ApJ...783...55C}; the top of the ovoid is associated with rapid wind velocities. 
We do not see these egg-like Alfvén lobes in the unscaled 
\Unscaled{5}{C1}AV~523
and
\Unscaled{D}{C2}V439~And
models.
We also observe some differences for the highly inclined current sheet stars between the scaled and the unscaled series of models. The Alfvén lobes of the unscaled models appear more rounded compared to the scaled models, which appear flattened near the current sheet. This flattening is sometimes accompanied by radial extrusions near the current sheet which gives the Alfvén lobe a `duck-billed' appearance; this is particularly prominent in the northern Alfvén lobe of \Scaled{6}{C1}AV~1693. 
Similar, but thicker extrusions appear in 
\Scaled{9}{C1}TYC~1987,
\Scaled{A}{C2}DX~Leo,
and
\Scaled{C}{C2}HH~Leo. The rapid decrease of the local Alfvén radius near the current sheet, and consequent flattening of the Alfvén surface lobe is also seen in e.g.\ \citet{2016A&A...594A..95A}; these are however not as pronounced as the 
\Scaled{6}{C1}AV~1693 case.

The average Alfvén radius \(R_\Alfven\) is the average radial distance from the stellar surface to the Alfvén surface. Considering the scalar-valued function \(\left|\vec r_\Alfven(\theta, \varphi)\right|\), the (radial) distance at which \(u\) first exceeds \(u_\Alfven\) for each point on the stellar surface \(S\), here parametrised by the polar angle \(\theta\) and the azimuth angle \(\varphi\), we get
\begin{equation}
    R_\Alfven = \frac{1}{4\pi R^2}\oint\nolimits_S \left|\vec r_\Alfven(\theta, \varphi) \right| \, \mathrm{d}S .
\end{equation}
The torque-averaged Alfvén distance is similarly the average value of 
\(\left|\vec r_\Alfven(\theta, \varphi)\times \uvec \Omega\right| \) 
over the stellar surface:
\begin{equation}
    \left|\vec r_\Alfven \times \uvec \Omega \right| 
    = \frac{1}{4\pi R^2}\oint\nolimits_S \left|\vec r_\Alfven (\theta, \varphi) \times \uvec \Omega \right| \,\mathrm{d} S
\end{equation}
The \(R_\Alfven\) value is a key parameter in one-dimensional models of stellar angular momentum loss \citep{1967ApJ...148..217W,1968MNRAS.138..359M,1984LNP...193...49M,1988ApJ...333..236K}. We find good agreement between the \(R_\Alfven\) of this work and the dipole scaling relations of \citet{2018ApJ...854...78F} based on polytropic models.

\begin{figure*}
    \centering
    \includegraphics{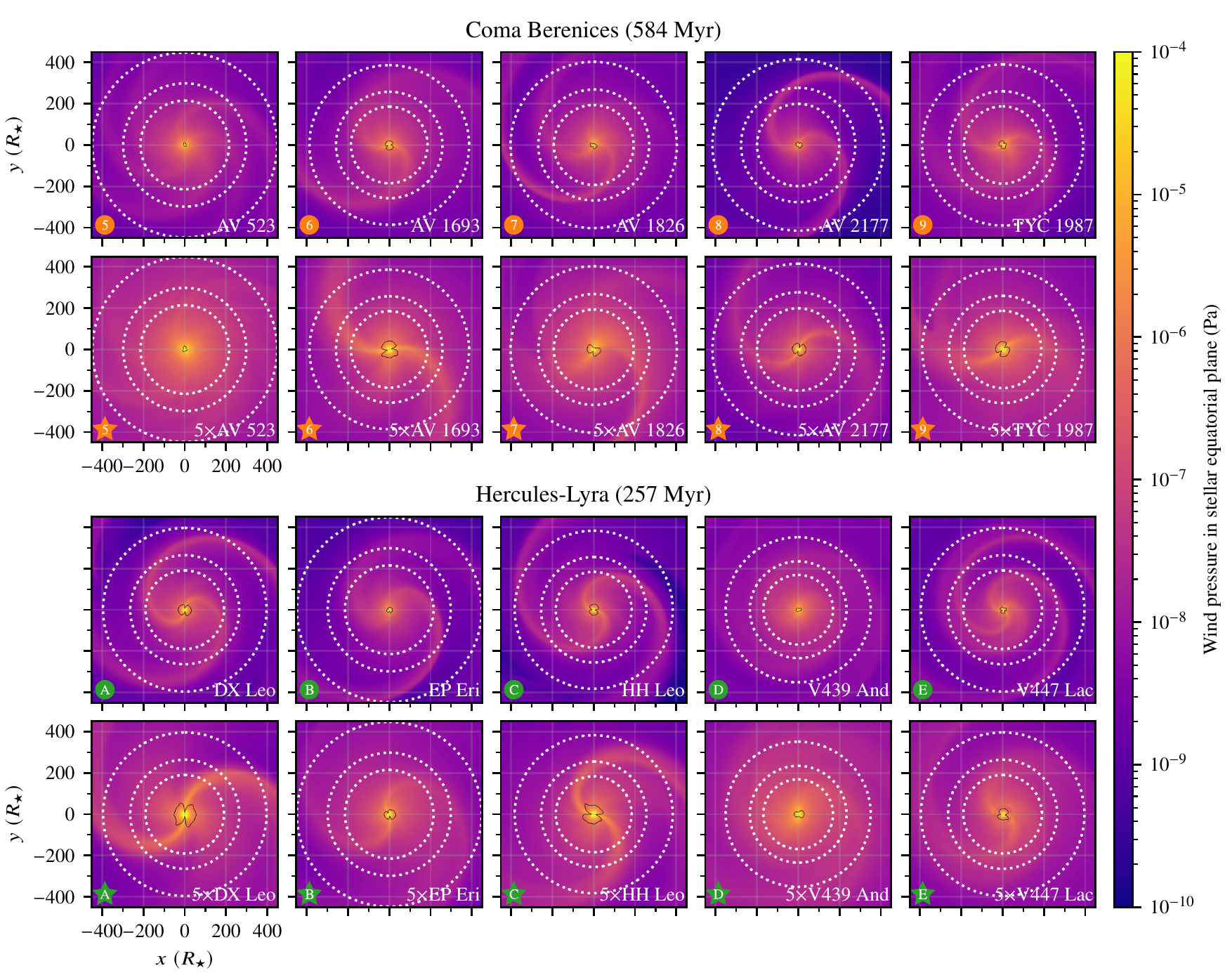}
    \caption{
        Wind pressure in the stellar equatorial plane. The order of the panels, which is the same as in Fig.~\ref{fig:paper2-br-fieldlines}, is indicated by the bottom left symbols in each panel. Co-rotating interacting regions (CIRs) produce multi-armed spiral structures; the number of arms depend on the geometry of the magnetic field near the stellar equator, and the amount of winding depends on the magnetic field strength and the stellar rate of rotation. Stars whose current sheet inclination is small give rise to less pronounced spiral structures. 
        The average distances of Venus-, Earth-, and Mars-like planets are indicated by white dotted circles. When visible at this scale, the intersection of the Alfvén surface and the stellar equatorial plane is indicated by a thin black closed curve. 
    }\label{fig:paper2-Pwind-IH-equatorial}
\end{figure*}
\subsection{Mass loss and angular momentum loss}\label{sec:mass-loss-angmom-loss}
The total wind mass loss \(\dot M\) and total wind angular momentum loss \(\dot J\) are two of the most studied quantities that may be derived from stellar and Solar wind maps. Mass loss values are difficult to constrain observationally, see the reviews of~\citet{2004LRSP....1....2W} and~\citep{2018haex.bookE..26V,2021LRSP...18....3V}.

From a flux argument it is clear that the mass loss is given by 
\begin{equation}
    \dot M = \oint\nolimits_S \rho \vec u \cdot \uvec n \,\mathrm{d} S
\end{equation}
where \(S\) is a closed surface surrounding the star and \(\vec u\) is the local wind velocity. 

The total wind angular momentum loss can be extrapolated from observed trends between stellar periods of rotation and stellar ages~\citep{1972ApJ...171..565S,2003ApJ...586..464B}. For a population of stars, average values of \(\dot J\) on timescales of millions of years and longer may be inferred. There does not, however, appear to be any way of observing \(\dot J\) for a particular star at a particular time, such as the \(\dot J\) values of the stars in this study. An argument based on angular momentum flux ~\citep[][see also \citetalias{paper1}]{1999stma.book.....M,2014MNRAS.438.1162V} gives the steady state \(\dot J\) values associated with our wind models,
\begin{equation}
    \label{eq:angmom_loss}
    \dot J 
    = \oint\nolimits_{S_\Alfven}
    \left(\uvec z \times \vec r \cdot \uvec n\right) \left( P+ \frac{B^2}{2 \mu_0 } \right)
    + ({\rho \vec V \cdot \uvec n}) \varpi^2 \Omega  
    \, {\rm d} S_\Alfven  
\end{equation}
where 
\(\vec \Omega\) is the star's angular velocity,
\(\varpi = \vec{r} - (\vec r \cdot \uvec z) \uvec{z}\),
\(\vec V = \vec u - \vec \Omega  \times \vec r\),
and 
\(S_\Alfven\) is the Alfvén surface. In our models we observe that the `effective corotation' term \(({\rho \vec V \cdot \uvec n}) \varpi^2 \Omega  \) dominates this expression at the Alfvén surface where we evaluate the integral. We note that \(\dot J\) may be evaluated over any closed surface enclosing the star, but this may involve the inclusion of additional terms in the integral, see \citet{2014MNRAS.438.1162V}. 
The presence of the angular velocity \(\Omega\) in the (dominant) effective corotation term of equation~\eqref{eq:angmom_loss} suggests that the parameter \(\dot J / \Omega\) may have a more direct link to the magnetic field strength and geometry; we return to this in Section~\ref{sec:Discussion}, where we find that this is indeed the case. 

\subsection{Wind pressure out to 1 au}\label{sec:wind-pressure}
Once the stellar winds become superalfvénic, discontinuities and shocks may form in the solution. 
As the stellar winds in our models travel outwards everywhere and
\(u_r \gg |\vec u_\perp| \), 
this typically occurs when a region of fast wind encounters a region of slower wind, giving rise to variations in the stellar wind properties such as the total wind pressure, which is shown in Fig.~\ref{fig:paper2-Pwind-IH-equatorial}. Note that discontinuities and shock may appear whenever \emph{relative} velocities exceed \(u_\Alfven\) in the solution, such as for the CMEs modelled by \citet{2020ApJ...895...47A}.

For a body moving through the stellar wind, such as an orbiting planet, the total wind pressure is the sum of the thermal pressure, magnetic pressure, and ram pressure,
\begin{equation}\label{eq:wind-pressure}
    P_\Wind= P+ |\vec B|^2 / (2 \mu_0) + \rho |\vec u+\vec v|^2
\end{equation}
\citep[see e.g.\@][]{2011MNRAS.414.1573V}, here \(\vec v\) is the planet's orbital velocity. The quantity \(P_\Wind\) is shown in Fig.~\ref{fig:paper2-Pwind-IH-equatorial} with \(\vec v=0\) i.e.\ neglecting the planet's orbital velocity. The planet's velocity may contribute significantly to \(P_\Wind\) for close-in exoplanets but for the would-be orbits of Venus, Earth and Mars (indicated by dotted white circles in Fig.~\ref{fig:paper2-Pwind-IH-equatorial}) we have \(|\vec u| \gg |\vec v|\). The Alfvén surface intersection with the \(xy\)-plane, outside of which the wind is superalfvénic, is shown as a thin black closed curve. Past a few stellar radii we observe that the wind pressure is dominated by the \(\rho |\vec u|^2\) term and that \(|\vec u|\approx u_r\).

In Fig.~\ref{fig:paper2-Pwind-IH-equatorial} we mainly observe two-, and three-armed structures of locally overdense wind, while the magnetic geometry of \Unscaled{5}{C1} AV~523 produces a five-armed structure. These structures arise when fast stellar wind encounters more slowly flowing wind originating from a different region of the stellar surface and are called co-rotating interacting regions (CIRs) \citep{1971JGR....76.3534B,1996ARA&A..34...35G}. 

We also calculate the average wind pressure for an Earth-like planet using equation~\eqref{eq:wind-pressure} and the Earth's orbital elements (see Table~\ref{tab:main-table}). 
The provided value is found by averaging \(P_\Wind\) over the stellar rotational phase and  the orbital phase of the Earth-like planet. As this parameter is evaluated near the stellar equator it is sensitive to variations in the current sheet inclination.

 A physical body such as a planet will give rise to a shock in the wind solution if the local wind is superalfvénic. For a magnetised planet the shock forms when the magnetic pressure generated by the planet is matched by the wind pressure. The so-called magnetospheric stand-off distance is then
\begin{equation}\label{eq:magnetospheric_standoff_distance}
    \left.R_\text{m} \middle/ R_\text{p}\right. = \left(B_0^2 \middle/ \left(2 \mu_0 \, \rho u^2\right)\right)^{1/6}.
\end{equation}
where \(R_\text{p}\) is the planet's radius and \(B_0\) is its dipolar magnetic field strength.
For a planet with the current-day magnetic field of the Earth, we use a dipolar magnetic field of \SI{0.7}{\gauss}; which includes factor of \(2\) of~\citet{1964JGR....69.1181M} accounting for currents in the magnetosphere. The resulting distances are given in Table~\ref{tab:main-table}.

\section{Discussion}\label{sec:Discussion}
In this section we examine trends in our results. Section~\ref{sec:effect-of-magnetic-scaling} considers the effect of magnetic scaling on the model results, and Section~\ref{sec:trends-and-correlations} considers correlations within a statistical framework. 

\subsection{Effect of magnetic scaling}\label{sec:effect-of-magnetic-scaling}
In this section we study the effect of magnetic scaling on the models in Table~\ref{tab:main-table} as well as the wind models of the Hyades stars 
    \Unscaled{0}{C0}Mel25-5,
    \Unscaled{1}{C0}Mel25-21,
    \Unscaled{2}{C0}Mel25-43,
    \Unscaled{3}{C0}Mel25-151, and
    \Unscaled{4}{C0}Mel25-179
and their scaled \(5B_\ZDI\) counterparts
from \citetalias{paper1}. 
The methodology is similar to the methodology of \citetalias{paper1} in that the two models we have for each star allows a direct investigation of the effect of the average magnetic field strength \(|B|\), and the residual variation caused by other factors such as magnetic geometry differences. 

\begin{figure}
    \centering
    \includegraphics{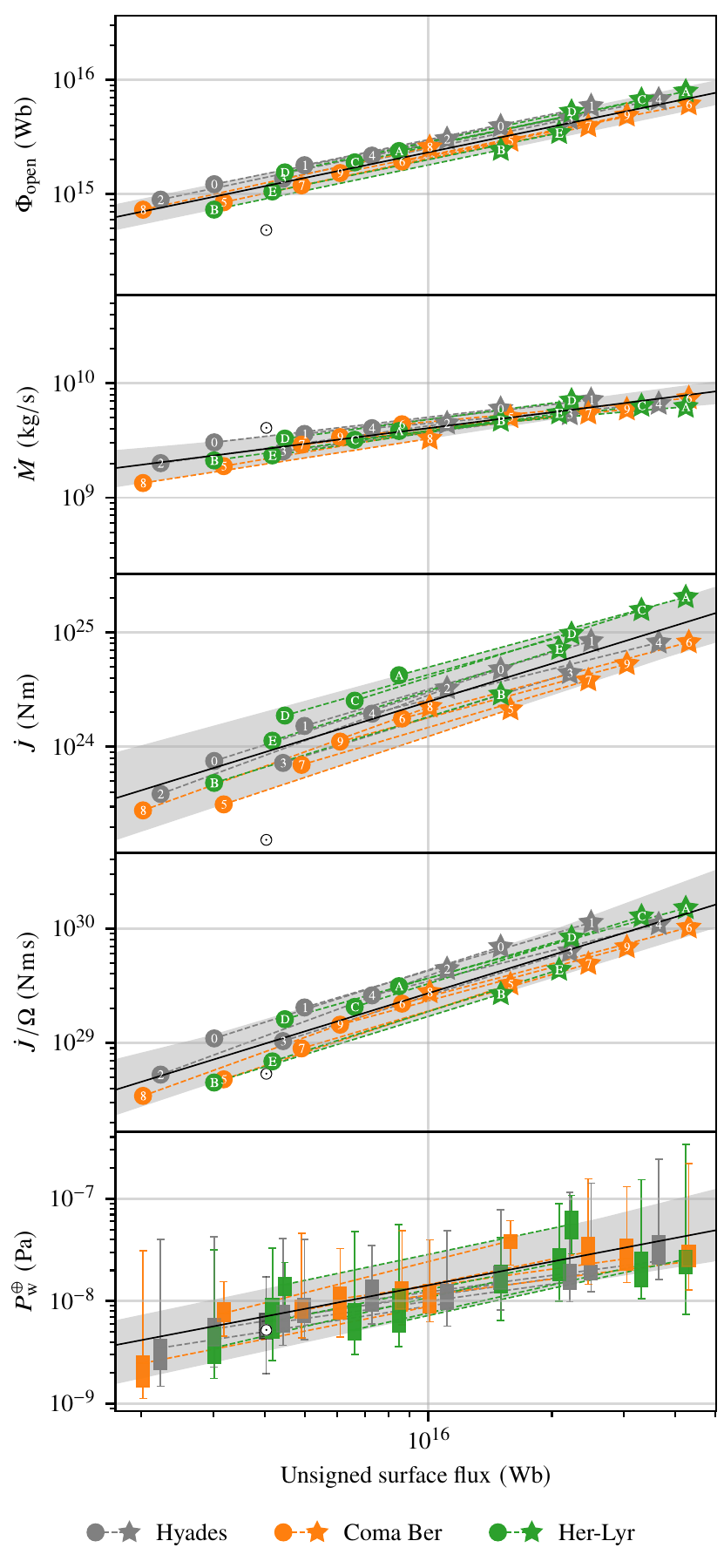}
    \caption{
        The effect of the magnetic scaling between the \(B_\ZDI\) and the \(5 B_\ZDI\) series and a lower bound on the residual variation.
        The open magnetic flux \(\Phi_\text{open}\), 
        wind mass loss rate \(\dot M\), 
        angular momentum loss rate \(\dot J\), 
        rotation-scaled angular momentum loss rate \(\dot J / \Omega\), 
        and wind pressure \(P_\Wind^\Earth\) 
        for an Earth-like planet is shown against the unsigned surface flux \(\Phi=4\pi R^2 |B_r|\). The variation in the wind pressure with stellar phase and planetary orbital phase is shown as boxplots. The shaded region represents the total variation in the dashed fitted barbell 
        power law lines, and the solid black power law line represents their midpoint.
        The \(y\) scale is the same in each panel, permitting a visual comparison of the slopes of the midpoint lines and the width of the residual variation between panels.
        The grey symbols and the Sun symbol represents the Hyades models and the Solar maximum wind model Sun-G2157 from \citetalias{paper1}.
        A more statistically rigorous approach to analysing the residual variation is given in Section~\ref{sec:trends-and-correlations} and Fig.~\ref{fig:trend-5x3-broken}.
    }\label{fig:orthogonal-trend-Unsigned-radial-flux-at-surface}
\end{figure}

By having two models for each star, that vary only in their absolute radial magnetic field strength, we are able to investigate the effect of the scaling of the magnetic field on the model results. As such these results are complementary to the relations found between stellar age and wind parameters such as  \citet{2015MNRAS.449.4117V} and \citet{2018ApJ...856...53P}, and the trends with age and rotation of \citet{2014MNRAS.441.2361V,2021LRSP...18....3V}.
The methodology of this section disentangles the link between age and rotation rate (although the stars are too young to adhere to the Skumanich law they are still spinning down; 
see \citet{2013A&A...556A..36G} and \citet{2018ApJ...862...90G}). The observed bimodality of the age-spin relationship \citep{2003ApJ...586..464B} for younger stars suggests that the younger stars in the sample may be in different regimes of magnetic braking depending on their periods of rotation; 
we do not, however, see any clear evidence of this in our models. A possibility is that our models lie below the critical angular velocity value for fast rotators, for which values from \SIrange{3}{15}{\Omega_\Sun} have been adopted 
\citep{1988ApJ...333..236K,1995ApJ...441..865C,2012ApJ...746...43R,2016A&A...587A.105A}. 
Fig.~\ref{fig:orthogonal-trend-Unsigned-radial-flux-at-surface} shows 
the open magnetic flux \(\Phi_\text{open}\),
the wind mass loss rate \(\dot M\), 
the angular momentum loss rate \(\dot J\),
the angular momentum loss rate scaled by the stellar rate of rotation \(\dot J / \Omega\), 
and the wind pressure for an Earth-like planet \(P_\Wind^\Earth\) plotted against the stellar
unsigned surface magnetic flux \(\SF = 4\pi R^2 |B_r|\). The values used in the plots are given in Table~\ref{tab:main-table}.
The variation in wind pressure with stellar rotation and orbital position (true anomaly) of the Earth-like planet is indicated by boxplots.

We have also found that the unsigned surface flux better predicts the model output values than the average surface radial field strength \(B_r\), hence we use the unsigned surface flux \(\SF\) as the independent variable in our fits.

For each star modelled, Fig.~\ref{fig:orthogonal-trend-Unsigned-radial-flux-at-surface} shows a barbell 
comprising a circle representing the star's \(B_\ZDI\) series value, a star symbol representing the star's \(5B_\ZDI\) series value (i.e.\ the symbols of Table~\ref{tab:main-table}), and a dashed line segment connecting the two.  
Each line segment has an equation of the form
\begin{subequations}
\begin{equation}
    \log_{10} \hat y_i(x) = \alpha_i \log_{10} x + \log_{10}\beta_i, \text{ so that } \hat y_i(x) \propto x^{\alpha_i}
\end{equation}
i.e.\ a straight line in a log-log plot and a power law in linear coordinates. The index \(i\) ranges over the stars in Table~\ref{tab:observed_quantities} and the barbell connects the model output for star \(i\) in the \(B_\ZDI\) with the model result for star \(i\) in the scaled \(5 B_\ZDI\) series. 
By comparing the unscaled \(B_\ZDI\) and the scaled \(5 B_\ZDI\) model for each star we can see the direct influence of the magnetic field strength on the model results. The geometric midpoint and range of the slope values \(\alpha_i\) are given in Table~\ref{tab:orthogonal_correlations}, and indicated as a black power law line \(\hat y(x) = \beta x^\alpha\) in Fig.~\ref{fig:orthogonal-trend-Unsigned-radial-flux-at-surface}. To indicate the amount of variation across each of the \(\hat y_i(x)\) curves, the area between the \(\min_i \hat y_i(x) \) and \(\max_i \hat y_i(x) \) curves is shaded in Fig.~\ref{fig:orthogonal-trend-Unsigned-radial-flux-at-surface}.

Table~\ref{tab:orthogonal_correlations} also gives a geometric measure of the residual variation around the  midpoint power law line; these \(\ymaxmingeom\) values represent the largest range of residuals in the fitted region \(X_\Star\) 
\begin{equation}
    \frac{ y_\text{max}}{ y_\text{min}} = \max_{x \in X_\Star} \,
    {
        \frac{\max \hat y_i(x)}{\min \hat y_i(x)}
    },
    \quad
    X_\Star = (\min \SF_\Star, \, \max \SF_\Star),
\end{equation}
i.e.\ the largest `height' of the shaded region in Fig.~\ref{fig:orthogonal-trend-Unsigned-radial-flux-at-surface}.
We also give the coefficients of determination (\(r^2\)-values) for the geometric midpoint \(\hat y(x)\) power law lines as they provide a measure of the quality of the fit without being influenced by the magnitude of the fitted \(\alpha\) coefficient. The \(r^2\) values are given by 
\begin{equation}
    r^2 = 1 - 
    \left.
        \sum{\big(\log_{10} y_j / \bar y \big)^2} 
    \middle/ \,
        \sum{\big(\log_{10} y_j / \hat y(x_j) \big)^2}
    \right.
    ;
\end{equation}
this is the standard  \citep[e.g.][]{1998ara..book.....D} definition of \(r^2\) subject to the logarithm identities \(\log y_j - \log \bar y = \log y_j/\bar y\) and \(\log y_j - \log \hat y(x_j) = \log y_j/\hat y(x_j)\), and where \(\bar y\) is the mean of the \(y\) values.
Note that in calculating the \(r^2\) values the \(j\) index ranges over the models in \emph{both} the scaled and the unscaled series (i.e.\ all the rows in Table~\ref{tab:main-table}), so that \(x_j\) and \(y_j\) represent the result of an individual model.
It bears repeating (see \citetalias{paper1}) that the quantities \(\alpha\), \(y_\text{max} / y_\text{min}\) and \(r^2\) are independent of any assumption about the statistical distribution of the model results. An analysis requiring some mild statistical assumptions is given in Section~\ref{sec:trends-and-correlations}.

\end{subequations}
\begin{table}
    \centering
    \caption{
        Clear trends from magnetic scaling and a lower bound on residual variation. 
        Power laws on the form \(\hat y \propto \Phi^\alpha\) are able to explain a large amount of the variation in the quantities of Table~\ref{tab:magnetic_averages} and Table~\ref{tab:main-table}. 
        The \(\alpha\) coefficients and their associated uncertainty indicate the midpoint of the power laws on the form \(\hat y_i(x) \propto \Phi^{\alpha_i}\) and the range of \(\alpha_i\) coefficients are indicated by the range on \(\alpha\). 
        A measure of the residual variation that cannot be attributed to variations in \(\SF\) is given by the \(\frac{y_\text{max}}{y_\text{min}}\) values. Aside from \(\dot J\), the largest residual variation is found for \(P_\Wind^\Earth\) and \(\dot J / \Omega\), suggesting that the magnetic geometry plays a large role in the determination of these quantities.
        The coefficients of determination (the \(r^2\) values) also show the residual variation after the fit to \(\Phi\) independently of the value of the exponent \(\alpha\), hence the two last rows have the same \(r^2\) value.
    }\label{tab:orthogonal_correlations}
    \sisetup{
        table-figures-decimal=3,
        table-figures-integer=2,
        table-figures-uncertainty=2,
        table-number-alignment=center,
        add-decimal-zero=true,
        add-integer-zero = false,
        omit-uncertainty = false
    }
    \input{tables/orthogonal_table.tex}
\end{table}

The middle panel of Fig.~\ref{fig:orthogonal-trend-Unsigned-radial-flux-at-surface} suggests decreasing trend in \(\dot J\) with stellar age between the 
Hercules-Lyra association (aged \SI{257 \pm 46}{\giga\year}) and the 
Coma Berenices cluster (aged \SI{584 \pm 10}{\giga\year}). 
The presence of the dominating effective corotation term \((\rho \vec V \cdot \uvec n) \varpi^2 \Omega\) in equation \eqref{eq:angmom_loss} describing the total angular momentum loss suggests that by scaling the angular momentum loss by the stellar rate of rotation will produce a tighter correlation with \(\SF\); we see that this is indeed the case in the fourth panel of Fig.~\ref{fig:orthogonal-trend-Unsigned-radial-flux-at-surface}, where \(\dot J / \Omega\) is plotted against \(\Phi\). 
The Hyades (aged \SI{625 \pm  50}{\mega\year}) wind models from \citetalias{paper1} lie between the Hercules-Lyra models and the Coma Berenices models when plotting \(\dot J\), and they are a bit higher than the average when plotting \(\dot J/\Omega\) against \(\Phi\). The overall spread is however reduced even when including the Hyades wind models. 

We briefly note some key features seen in Table~\ref{tab:orthogonal_correlations} and Fig.~\ref{fig:orthogonal-trend-Unsigned-radial-flux-at-surface}:
\begin{enumerate}
    \item We see that \(|B_r|\) and \(\max |B_r|\) are proportional to \(\Phi\) and as such have \(\alpha=1\); this is enforced by the model and methodology as, for each star, the relation between \(\Phi\), \(|B_r|\) and \(\max |B_r|\) are linear. The variation in these parameters are caused by the variation in the magnetogram geometry and quantified by \({y_\text{max}}/{y_\text{min}}\) in Table~\ref{tab:orthogonal_correlations}. 
    \item The average surface magnetic field strength \(|\vec B|\) is closely correlated with \(\Phi\) as well, indicating that the radial magnetic field, rather than rotation effects determine the non-radial components of \(\vec B\). 
    \item As in \citetalias{paper1} the Alfvén radius \(R_\Alfven\) and the torque-averaged Alfvén radius exhibit very similar correlation coefficients \(\alpha\approx\num{0.39}\) and similar variation measures and coefficients of determination. 
    \item With a larger number of stars (15 vs. 5) in comparison to \citetalias{paper1} we expect to see larger variation and higher values of the variation measure as it is a lower bound on the population variation and can only increase with increasing numbers of stellar models being included. We do observe larger variations in \(\Phi_\text{open}\), \(\dot M\), \(\dot J\), \(P_\Wind^\Earth\) and \(R_\text{m}\) compared to \citetalias{paper1}. 
    \item The scaled angular momentum \(\dot J/\Omega\) exhibits a smaller variation \({y_\text{max}}/{y_\text{min}}\) than \(\dot J\) itself as was seen in Fig.~\ref{fig:orthogonal-trend-Unsigned-radial-flux-at-surface}; the variation measure is 
    \SI{311}{\percent} and \SI{568}{\percent} respectively. 
    \item The inclusion of \(\dot J/\Omega\) means that the wind pressure for an Earth-like planet \(P_\Wind^\Earth\) is the parameter that is least well predicted by \(\Phi\).
\end{enumerate}

\subsection{Statistical trends and correlations}\label{sec:trends-and-correlations}
In this section we apply an ordinary least square (OLS) fit to the models of the \(B_\ZDI\) and \(5B_\ZDI\) series. This approach is complementary to the approach in Section~\ref{sec:effect-of-magnetic-scaling} where the effect of the magnetic scaling on each star model was investigated. The ordinary least square fit may be less geometrically intuitive than the effect of magnetic scaling, but the powerful statistical machinery of OLS does permit a more quantitative approach. By log-transforming our data they satisfy the assumptions required by OLS \citep[see e.g.][]{1998ara..book.....D} including homoscedasticity and normality.
As well as the trends themselves, OLS permits a quantitative analysis of the residual variation in the dataset, expected to correspond to variations in magnetic geometry. 

\begin{figure*}
    \centering
    \includegraphics{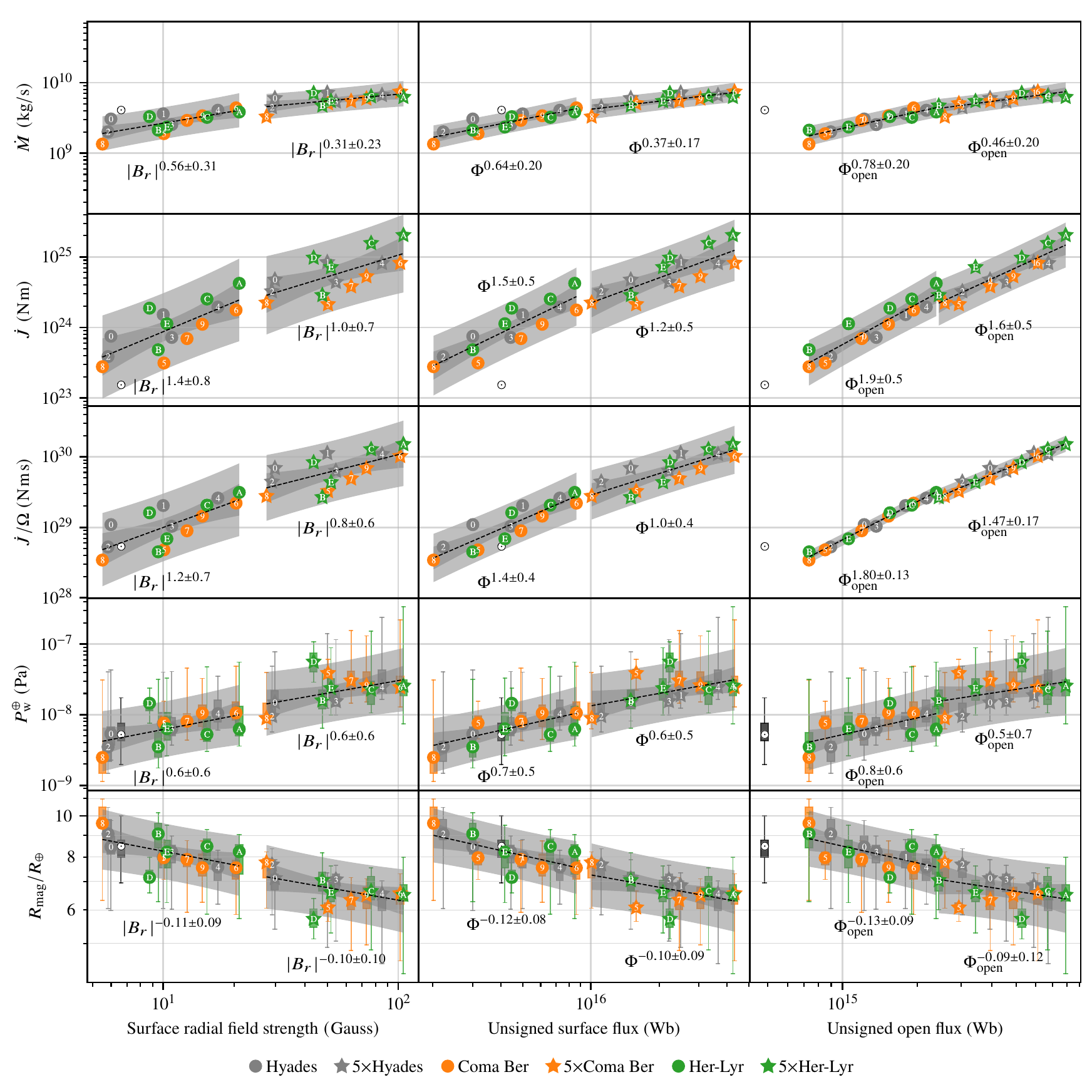}
    \caption{
        Trend lines and variation bands for (from top to bottom)
        mass loss, 
        angular momentum loss, 
        scaled angular momentum loss, 
        wind pressure at \SI{1}{\astronomicalunit}, and 
        magnetospheric stand-off distance for an Earth-like planet plotted against (from left to right)
        the surface radial magnetic field strength, 
        the unsigned surface flux,
        and the unsigned open flux.
        The symbols used in the plot corresponds to the tabulated values in Table~\ref{tab:main-table}.
        The dashed black lines are power law fits made to the \(B_\ZDI\) series and the \(5B_\ZDI\) series, and the dark grey regions represent confidence bands calculated at \SI{95}{\percent} confidence level. The light grey regions represent prediction bands; subject to the OLS assumptions, there is a \SI{95}{\percent} chance that further wind models lie inside the shaded region. Key values behind this plot are tabulated in Table~\ref{tab:correlations-split}.
        The Hyades wind models from \citetalias{paper1} (grey symbols) are included in the fits. The Solar maximum wind model Sun-G2157 from \citetalias{paper1} is included in the plots as a Sun symbol; this model is not included in the fits. The Solar wind parameters vary on many timescales \citep[see e.g.\ ][]{1998ApJ...498L.165W,2018ApJ...864..125F} so the Sun-G2157 model reflects only a snapshot in time.
        In the two bottom rows the point-to-point variation is indicated by boxplots.        
    }\label{fig:trend-5x3-broken}
\end{figure*}

Figure~\ref{fig:trend-5x3-broken} shows trend lines (dashed black lines), confidence bands (dark grey regions) and prediction bands (light grey regions) for the 
mass loss, 
angular momentum loss, 
scaled angular momentum loss, 
wind pressure at \SI{1}{\astronomicalunit}, and 
magnetospheric stand-off distance for an Earth-like planet 
plotted against 
the surface radial magnetic field strength, 
the unsigned surface flux,
and the unsigned open flux.
Select statistical parameters relating to Fig.~\ref{fig:trend-5x3-broken} are given in Table~\ref{tab:correlations-split}. We provide the \(a\) parameter with a \SI{95}{\percent} confidence interval, the coefficients of determination, the \(p\)-values, and a measure \(\ymaxminstat\) of the mean size of the prediction interval at the data points of the \(B_\ZDI\) series and the \(5B_\ZDI\) series; this value represents a measure of the height of the light grey shaded prediction band region in Fig.~\ref{fig:trend-5x3-broken}.  The reported statistical parameters are calculated in the standard way \citep[see e.g.][]{1998ara..book.....D} using the 
\textsc{statsmodels} Python package \citep[version 0.12.1,][]{seabold2010statsmodels}. 
In comparison to the shaded residual variation in 
Fig.~\ref{fig:orthogonal-trend-Unsigned-radial-flux-at-surface} 
and the \(\ymaxmingeom\) values in 
Table~\ref{tab:orthogonal_correlations}, 
the light grey shaded prediction bands in Fig.~\ref{fig:trend-5x3-broken} are wider and the \(\ymaxminstat\) values (Table~\ref{tab:correlations-split}) are greater. 
The bands in Fig.~\ref{fig:orthogonal-trend-Unsigned-radial-flux-at-surface} are a geometric measure of the minimum variation and, as such, they are a lower bound on the variation attributed to differences in magnetic geometry. 
Prediction bands, such as the light grey \SI{95}{\percent} prediction bands in Fig.~\ref{fig:trend-5x3-broken}, come with the different guarantee that, as long as the OLS assumptions are not violated, \SI{95}{\percent} of new wind models with similar parameters to the ones in this study would fall inside the light grey shaded region.
\begin{table*}
    \centering
    \caption{
        Numerical summary of the fitted curves in Fig.~\ref{fig:trend-5x3-broken}. This table shows the log-log correlations between quantities of Tables~\ref{tab:magnetic_averages} and \ref{tab:main-table} and the average unsigned radial field strength \(|B_r|\), the unsigned surface magnetic flux \(\Phi\),  and the unsigned open magnetic flux \(\Phi_\text{open}\). The \(a\) values columns give the exponent of the fitted power laws with \SI{95}{\percent} confidence intervals. The coefficients of determination are given in the \(r^2\) columns. The \(\ymaxminstat\) columns give the average width of the \SI{95}{\percent} prediction intervals; the average is taken over the data points of the series. The probability values (\(p\)-values) are given in the eponymous columns.
    }\label{tab:correlations-split}
    \sisetup{
        table-figures-decimal=2,
        table-figures-integer=1,
        table-figures-uncertainty=2,
        table-number-alignment=center,
        add-decimal-zero=false,
        add-integer-zero = false,
        omit-uncertainty = false
    }
    \input{tables/correlations-rvalues-pvalues-predint-1-5.tex}
\end{table*}

The dashed black trend lines on Fig.~\ref{fig:trend-5x3-broken} are power laws on the form \(y(x)\propto x^{a}\); they are obtained from fitting a curve on the form
\begin{equation}\label{eq:loglog-fit}
    y(x) = b x^{a} \text{ so that } \log_{10} y(x) = \log_{10} b + a \log_{10} x
\end{equation}
to the set of model values. The scaling laws that are overprinted such as \(|B_r|^{0.56\pm0.31}\) in the top left panel indicate that the true value of \(a\) lies within \(0.56\pm0.31\) with \SI{95}{\percent} probability, so that \(\dot M \propto |B_r|^{0.56\pm0.31} \) for the data in the \(B_\ZDI\) series of models. 

In Fig.~\ref{fig:trend-5x3-broken} the true power law is \SI{95}{\percent} likely to lie within the dark grey confidence band region surrounding the line of best fit, i.e.\ confidence bounds on the entire fitted curve on the form \(y(x)=bx^a\). Furthermore, new observations are \SI{95}{\percent} likely to lie within the wider light grey prediction band region. As such the height of the confidence bands represents a confidence interval on the residual variation caused by differences in the magnetic geometry and variations in the fundamental parameters of Table~\ref{tab:observed_quantities}. The variance of this prediction is the sum of the variance of the fitted curve \(y(x)=bx^a\) and the mean square error of the regression, hence the light grey prediction bands are always wider than the dark grey confidence bands. 

As in Fig.~\ref{fig:orthogonal-trend-Unsigned-radial-flux-at-surface}, we have included the wind models of the five Hyades stars from \citetalias{paper1} as well as the models of the ten stars of this work. 
In comparison to the similar plot in \citetalias{paper1}, the inclusion of all fifteen wind models in the \(B_\ZDI\) and the \(5B_\ZDI\) significantly reduces the width of the \SI{95}{\percent} dark grey confidence bands around the fitted curves \(y(x)\) in Fig.~\ref{fig:trend-5x3-broken}.
Considering the correlations with \(|B_r|\) (the leftmost column of panels in Fig.~\ref{fig:trend-5x3-broken}), we see significant correlations for \(\dot M\), \(\dot J\), \(\dot J/\Omega\) in both the \(B_\ZDI\) model series and the \(5B_\ZDI\) model series.

With the exception of \(P_\Wind^\Earth\) and \(R_\text{m}\) we find \(p\leq 0.01\) for all the correlations under consideration. For \(P_\Wind^\Earth\) and \(R_\text{m}\) we observe \(p\leq 0.05\) except for \(P_\Wind^\Earth (\Phi_\text{open})\) and \(R_\text{m} (\Phi_\text{open})\) in the \(5B_\ZDI\) series (bottom right of Table~\ref{tab:correlations-split}). The comparatively large amount of unexplained variation in \(P_\Wind^\Earth\) and \(R_\text{m}\) indicate that other parameters than \(|B_r|\), \(\SF\) and \(\Phi_\text{open}\) play a large role in determining \(P_\Wind^\Earth\) and \(R_\text{m}\).

In Fig.~\ref{fig:trend-5x3-broken} we have also included the Solar maximum value from \citetalias{paper1} (indicated by a Sun symbol) in each panel as it falls inside the range of \(|B_r|\) values modelled in this work. The Sun model is excluded from regression analysis; it is included here for ease of comparison with the plots of \citetalias{paper1}.
    
The middle column in Fig.~\ref{fig:trend-5x3-broken} shows correlations with the surface flux \(\Phi\) (recall that \(\Phi=4\pi R^2 |B_r|\)). The most notable difference between the correlations with \(|B_r|\) (leftmost column) is that the light grey prediction bands are tighter in the correlations with \(\Phi\) i.e.\ there is a smaller amount of unaccounted for variation when using the unsigned surface flux as the independent variable of the fit. 

We also include the correlations with the open flux \(\Phi_\text{open}\) in the rightmost column of Fig.~\ref{fig:trend-5x3-broken}. The light grey prediction bands are the narrowest for \(\Phi_\text{open}\), strikingly so for \(\dot J/\Omega\). 
This result indicates a very tight correlation between \(\dot J/\Omega\) and \(\Phi_\text{open}\) produced by the \awsom{} model.    
Tight correlations between \(\Phi_\text{open}\) and \(\dot J\) and \(\dot M\) are the basis for parametric methods of estimating these parameters, see \citet{2015ApJ...798..116R}.
It bears repeating here that \(\Phi_\text{open}\) is itself a model output so that this rightmost column of Fig.~\ref{fig:trend-5x3-broken} is a comparison between two model outputs that cannot be known without first running the models to completion. The complementary approach of a potential extrapolation \citep{1969SoPh....6..442S,1969SoPh....9..131A,1984PhDT.........5H,1992ApJ...392..310W} of the coronal magnetic field can, however, be used to estimate \(\Phi_\text{open}\), subject to the parameters of this method. 

The parameter \(\dot J/\Omega\), which was introduced to reduce the residual variation due to differences in stellar period of rotation, exhibits reduced residual variation compared to \(\dot J\) for all the independent variables used in Fig.~\ref{fig:trend-5x3-broken}: \(|B_r|\), \(\Phi\), and \(\Phi_\text{open}\). The \(a\) values found for \(\dot J/\Omega\) are smaller than the ones found for \(\dot J\), but they lie inside each other's \SI{95}{\percent} confidence intervals. The light grey prediction bands are wider for \(\dot J\) than for \(\dot J/\Omega\); numerically the width of the prediction bands may be compared by checking the \(\ymaxminstat\) column in Table~\ref{tab:correlations-split}.

Comparing the slopes of curves fitted to the \(B_\ZDI\) and the \(5B_\ZDI\) series in each panel, we observe that the fitted exponents to the two series are mostly compatible in the sense that the mean \(a\) value for the \(B_\ZDI\) series lies inside the \SI{95}{\percent} confidence interval of \(a\) for the the \(5B_\ZDI\) series, and vice versa. The exceptions are the wind mass loss parameter where there are some signs of saturation occurring at higher surface magnetic field strengths. Similar signs of saturation is seen in the \awsom{} wind models of \citet{2016A&A...594A..95A} and \citet{2018ApJ...856...53P} as we observed in \citetalias{paper1}; see also Section~\ref{sec:scaling-law-comparison}. 

We now focus on the lower part of Fig.~\ref{fig:trend-5x3-broken} where we plot the correlation between the wind pressure at \SI{1}{\astronomicalunit} and the resulting magnetospheric stand-off distance for an Earth-like planet \(R_\text{m}\).  While the curves of best fit have noticeable slopes for \(|B_r|\), \(\Phi\), and \(\Phi_\text{open}\), these correlations lie around the threshold of statistical significance, indicating that \(|B_r|\) and \(\Phi\) do not have strong predictive power over these values. For the Hercules-Lyra stars there appears to be a correlation between the residual and the current sheet inclination;
\Unscaled{D}{C2}V439~And has a low inclination while 
\Unscaled{A}{C2}DX~Leo,         
\Unscaled{B}{C2}EP~Eri, and          
\Unscaled{C}{C2}HH~Leo are highly inclined.
We also see this pattern for the low inclination  
\Unscaled{5}{C1}AV~523         
and the high inclination
\Unscaled{8}{C1}AV~2177. These patterns appear in both the \(B_\ZDI\) series and the \(5B_\ZDI\) series of models.         

In \citetalias{paper1} we observed a strong correlation between \(\Phi_\text{open}\) and \(P_\Wind^\Earth\). The inclusion of the wind models from this work suggest that the tightness of this particular correlation was spurious; we now see little difference in explanatory power between the fitted curves \(P_\Wind^\Earth (|B_r|)\), \(P_\Wind^\Earth (\Phi)\), and \(P_\Wind^\Earth (\Phi_\text{open})\) in Fig.~\ref{fig:trend-5x3-broken}.

\subsection{Comparison with known scaling laws}\label{sec:scaling-law-comparison}

Many observation-based and semi-empirical scaling relations have been put forward in the literature of mass loss and angular momentum loss against age and rotation rates.
    
Projecting backwards in time, \citet{2002ApJ...574..412W, 2005ApJ...628L.143W} predict mass loss values increasing from Solar values to \(\mysim 10^2 \dot M_\odot\) at a stellar age of \SI{0.7}{\giga\year}, followed by lower mass losses for even younger stars; this is the so-called `wind dividing line'. In our models we do not find such a threshold but we may see some signs of saturation in \(\dot M\). Recently it has been suggested that the wind dividing line might not exist as its existence is conjectured based on a few data points only; see e.g.\ the review of \citet{2021LRSP...18....3V}. 

The modelling study of \citet{2013PASJ...65...98S} found \(\dot M \propto t^{-1.23}\) and no saturation of \(\dot M\) for very young stars; the model predicts mass loss values of \(\mysim\num{2e2} \dot M_\Sun\) for the Hyades and Coma Berenices, and of \(\mysim\num{5e2} \dot M_\Sun\) at the age of the Hercules-Lyra association. 

\begin{figure}
    \centering
    \input{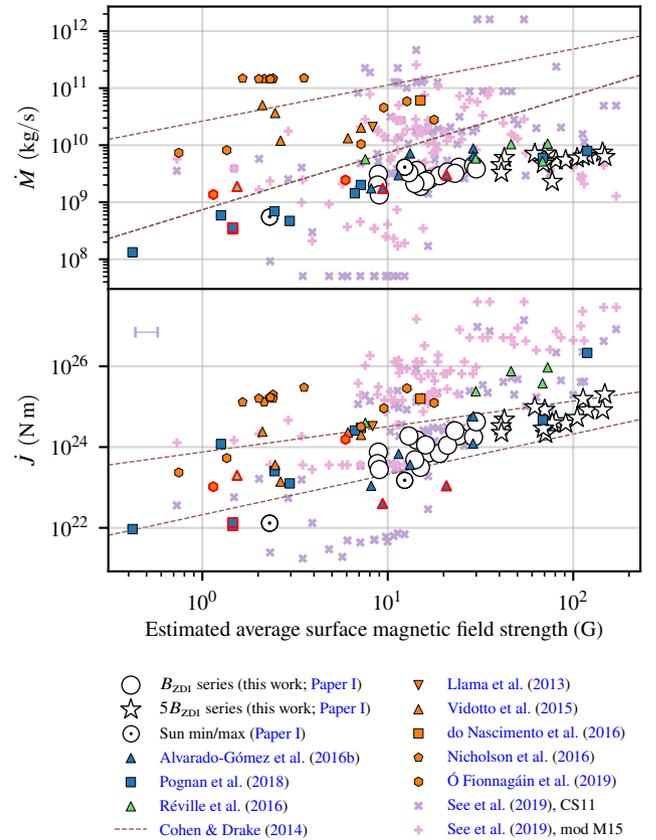}
    \caption{ 
        Comparison of wind mass \(\dot M\) and angular momentum loss \(\dot J\) values of this study and \citetalias{paper1} with literature values. The white Sun symbols, circles, and stars represent wind model results from this work and \citetalias{paper1}. The black and red outlined symbols represent three-dimensional MHD simulations of stars and the Sun respectively. The colour of these symbols represent the model type: blue for \awsom{}, orange for ideal MHD, and green for the \pluto{} model. The symbols with no outline represent populations of stars to which scaling laws have been applied; for these the interval in the upper left corner of the \(\dot J\) panel indicates the uncertainty in our estimates of the \(|B|\) values. When clusters of data points with identical \(\dot M\) values are seen they refer to the same star at different epochs. The brown dashed lines refer to the scaling laws of \citet{2014ApJ...783...55C} with Sun-like coronal densities 
        \numrange{2e14}{2e15} particles per cubic meter and a rotational period of \SI{10}{\day}. CS11 and mod M15 refers to different methods used by \citet{2019ApJ...886..120S}. A detailed description of many aspects of this plot is available in \citetalias{paper1}. 
    }\label{fig:results-context}
\end{figure}

These values are significantly higher than the mass loss values found from MHD studies similar to our own.
In \citetalias{paper1} we compared our stellar wind models to a large sample of literature values including 
the \awsom{} based models of  \citep{2016A&A...594A..95A},  %
and the hot corona ideal MHD models of
\citet{2013MNRAS.436.2179L,2015MNRAS.449.4117V,2016MNRAS.459.1907N,2016ApJ...820L..15D,2019MNRAS.483..873O}. 
and the hybrid models of \citep{2018ApJ...856...53P},  %
as well as the \pluto{} code MHD models of \citep{2016ApJ...832..145R}.  %
The mass loss and angular momentum loss found in these studies are compared to our own in Figure~\ref{fig:results-context}. 

We also include a comparison with the scaling laws of \citet{2014ApJ...783...55C} for two different coronal number densities of \(n = \SI{2e14}{\per\cubic\meter}\) and \(n = \SI{2e15}{\per\cubic\meter}\), both with a period of rotation of 10 days. Decreasing the period of rotation shifts the curve upwards. 
The stellar populations of \citet{2017MNRAS.466.1542S,2019ApJ...886..120S} are also included in Figure~\ref{fig:results-context}. For the differences between the CS11 and mod~M15 methodologies we refer the reader to the \citet{2019ApJ...886..120S} and the short discussion in \citetalias{paper1}. Due to inconsistent reporting of magnetic quantities between different studies, the quantity \(|B|\) has in some cases been estimated from other parameters; the methodology applied is described in \citetalias{paper1}.

From Figure~\ref{fig:results-context} it is clear that our model results show good comparison with the \awsom{} based models of \citep{2016A&A...594A..95A}, the hybrid models of \citep{2018ApJ...856...53P} and the \pluto{} code models of \citep{2016ApJ...832..145R}, particularly for \(\dot M\). The ideal MHD models and the populations of \citep{2017MNRAS.466.1542S,2019ApJ...886..120S}, on the other hand, predict significantly higher values of both \(\dot M\) and \(\dot J\).

We note here that some ideal MHD studies such as \citet{2019MNRAS.483..873O} also find evidence of saturation in \(\dot J\) as a function of rotational angular velocity \(\Omega\), but not necessarily in \(\dot J\) as a function of \(\vec B\).
Compared to \citetalias{paper1} we essentially see a continuation of the trends observed for the Hyades, as previously discussed in Section~\ref{sec:effect-of-magnetic-scaling}--\ref{sec:trends-and-correlations}. \\ \\

For wind pressure and magnetospheric stand-off distance estimates, a wide range of values have been reported in the literature. Comparing to three recent studies, we find  our \(P_\Wind^\Earth\) and \(R_\text{m}/R_\text{p}\) values occupying a middle range. 
Studying radio emissions of close-in exoplanets,
\cite{2015MNRAS.450.4323S} %
modelled the wind ram pressure of the stars HD184733, HD179949, and \(\tau\) Boötis, 
at orbital distances of \SIrange{0.03}{0.05}{\astronomicalunit} and found wind ram pressures of 
\SI{\sim 1.4}{\micro\pascal}
\SI{\sim 0.33}{\micro\pascal} and
\SI{\sim 0.23}{\micro\pascal}
for rotation periods of  
\SI{12.5}{\day}, 
\SI{7.6}{\day}, and
\SI{3.31}{\day} respectively 
using a potential field extrapolation of the coronal magnetic field and the Wang-Sheeley-Arge \citep{1990ApJ...355..726W,2000JGR...10510465A} method for determining wind velocity. At a similar close-in distance, our models give wind pressure values around \SI{1}{\micro\pascal} for the \(B_\ZDI\) series, and \SI{10}{\micro\pascal} for the \(5B_\ZDI\) series; our models thus give nearly an order of magnitude higher pressure values. 

In their Solar wind in time study,
\citet{2018MNRAS.476.2465O} %
derived two scaling laws: one for wind ram pressure at \SI{1}{\astronomicalunit} as a function of age, and one for wind pressure as a function of rotational angular velocity.
Their models incorporate a break at around \(1.4\Omega_\Sun\) or \SI{2}{\giga\year}, before which the magnetospheric stand-off distance \(R_\text{m} \propto \Omega^{-0.32}\); the model predicts \(R_\text{m}/R_\text{p}\) values of 3--5 for rotational periods of \SIrange{5}{12}{\day}. The detailed MHD models Sun-Earth wind-planet interactions by \citet{2019MNRAS.489.5784C} used  \(P_\text{ram} \propto \Omega^{0.27}\) for the wind ram pressure, and also found \(R_\text{m}\) values \(R_\text{m}/R_\text{p}\) values of 3--5 for rotational periods in the range \SIrange{3}{13}{\day}. Both these value ranges are smaller than our magnetospheric stand-off distance \(R_\text{m}/R_\text{p}\) range of 6--10, and predict wind pressure \(P_\Wind\) values \(\mysim50\) times greater than ours, given the \(R_\text{m}\propto P_\Wind^{-1/6}\) relation of equation~\eqref{eq:magnetospheric_standoff_distance}.
 
Although the comparisons made here are rudimentary, it seems clear that a wide range of wind pressure values have been reported in the literature. The large range of values is a consequence of the uncertainties surrounding \(\dot M\) (see Figure~\ref{fig:results-context}) as the ram pressure \(P_\text{ram} \propto \rho u^2\) and the wind mass loss \(\dot M \propto \rho u\) depend on the same hard-to-characterise quantities of wind density \(\rho\) and speed \(u\). The detection of radio emissions from close-in exoplanets could potentially help constrain \(\dot M\) and \(P_\Wind\) in the future \citep[see e.g.\ ][]{2019MNRAS.483..873O,2021MNRAS.504.1511K}.

\section{Conclusions}\label{sec:Conclusions}
We have modelled the winds of ten young, Solar-type stars in the Hercules-Lyra association and the Coma Berenices cluster aged 
\SI{\mysim 0.26}{\giga\year}
and 
\SI{\mysim 0.58}{\giga\year}
respectively.
By driving a state-of-the-art Solar system code \batsrus{}/\awsom{}, we obtain fully three dimensional wind maps of the stars' coronae and inner astrospheres. 
To account for the uncertain average magnetic field strength in ZDI \citep[e.g.\ ][]{2019MNRAS.483.5246L}, we create a second series of wind models, the \(5B_\ZDI\) series where the surface magnetic field strength is increased by a factor of 5, so that the magnetic energy is 25 times greater than in the unscaled \(B_\ZDI\) models. 
Combining these results with our previous models of similar stars in the Hyades
(aged \SI{\mysim 0.63}{\giga\year}), published in \citetalias{paper1}, gives a large sample of \(2\times15\) wind models. 

The complicated surface magnetic field geometries all give rise to dipole-like coronal magnetic fields and two-lobed Alfvén surfaces. The inclination of the dipole-like coronal magnetic fields take on a range of values seemingly in the full \SIrange{0}{90}{\degree} range. The \(5B_\ZDI\) series of wind models give rise to larger, more complex Alfvén surfaces, larger wind speeds, and larger regions of closed magnetic field lines.

The effect of the magnetic scaling between pairs of stellar wind models in the \(B_\ZDI\) and \(5B_\ZDI\) series in Section~\ref{sec:effect-of-magnetic-scaling} shows positive correlations with the surface magnetic flux for all considered parameters except, as expected, for the magnetospheric stand-off distance 
\(R_\text{m} \propto (P_\Wind^\Earth)^{-1/6}\).
The largest amount of residual variation is found for \(\dot J\) and the second largest amount of residual variation is found for \(P_\Wind^\Earth\). By controlling for the stellar rate of rotation in the \(\dot J/\Omega\) parameter the largest amount of residual variation is \(P_\Wind^\Earth\) where the inclination of the magnetic axis appears to play a role.

The conclusions of the geometric analysis of the effect of magnetic scaling in Section~\ref{sec:effect-of-magnetic-scaling} is strengthened by the complementary statistical analysis of the \(B_\ZDI\) and \(5B_\ZDI\) series as separate populations in Section~\ref{sec:trends-and-correlations}. Here we find \SI{95}{\percent} confidence intervals on the predicted total range of variation of wind models of young Solar-type stars in our age range of \SIrange{\mysim 0.26}{\mysim 0.63}{\giga\year}. The intervals correspond to a variation of \(\mysim 2\) in \(\dot M\) values and \(\mysim 4.5\) in \(\dot J / \Omega\) values. The large effect of the stellar rate of rotation is evident as the variation in \(\dot J\) itself is a factor of 6--7. 

There appears to be growing \citep{2016ApJ...833L...4G,2018PNAS..115..260D,2021ApJ...916...96A,2021MNRAS.504.1511K,2021MNRAS.500.3438O} interest in scaling the Poynting flux-to-field ratio \(\Pi_\Alfven/B\); variations in this parameter affect the steady-state wind mass loss as \(\dot M\) is roughly proportional to \(\Pi_\Alfven/B\) \citep{2020A&A...635A.178B}. \citet{2021ApJ...916...96A} scaled \(\Pi_\Alfven/B\) by a factor of 27 when modelling the young Solar-type star \(\kappa^1\) Ceti based on far UV observations. In this work where we use the Solar \(\Pi_\Alfven/B\) value and considered a scaled \(5B_\ZDI\) magnetic field, thus the mass loss rates do not go as high as in ideal MHD models. 
If a scaling of 27 or more of the Poynting flux is realistic for young, Solar-type stars, it means that many wind parameters of this work are underestimated by an order of magnitude. This could close the gap between ideal MHD models and the \awsom{} wind models observed in Section~\ref{sec:scaling-law-comparison}. More research on constraining the value of this parameter for young Solar-type stars would therefore be of interest.

We do not observe any sign of the `wind dividing line' of \citep{2004AdSpR..34...66W,2005ApJ...628L.143W} in our wind models, this is expected as the effect of photospheric currents and the resulting toroidal magnetic field is not part of the wind model input. The toroidal magnetic field is not expected to influence the steady state wind, but the toroidal magnetic energy would serve as an energy reservoir for transient events such as flares and coronal mass ejections \citep{2013MNRAS.431..528J}.

We do not observe any clear trends between the effective magnetogram degree estimates \(\ell_{.90}\) and \(\ell_{.99}\) in 
Section~\ref{sec:coronal_structure}
and the model derived quantities in 
Section~\ref{sec:Discussion}.
The large variations in wind parameters found by \citet{2015ApJ...807L...6G} for pure quadrupolar and higher degree fields do not seem to appear easily when driving the model with ZDI-derived mixed degree surface magnetic fields.  

The wind models in this work are in weak agreement with the studies of 
\citet{2018ApJ...854...78F} 
and 
\citet{2019ApJ...886..120S} 
which suggested that the dipolar magnetic field component dominates the wind angular momentum loss when wind mass loss does not exceed a threshold value; the mild saturation we observe in mass loss with increasing surface magnetic field strength does however mean that all the wind models in this work lay below the threshold value of mass loss.

It was noted by \citep{2018ApJ...864..125F} that two-temperature MHD models such as \batsrus{}/\awsom{}, that recover the bimodality of the fast and slow solar wind, have not yet been used to formulate scaling relations. With the set of \(2 \times 15\) wind models and resulting scaling laws provided here, this work may represent a small contribution to addressing this issue.

\section*{Acknowledgements}
    DE is funded by a University of Southern Queensland 
    (USQ) International Stipend Research Scholarship (ISRS) and a
    USQ International Fees Research Scholarship (IFRS).
    This research was undertaken using the University of Southern Queensland (USQ) Fawkes HPC which is co-sponsored by the Queensland Cyber Infrastructure Foundation (QCIF), see~\url{www.usq.edu.au/hpc}. 
    This work has made use of the \href{http://vald.astro.uu.se/}{Vienna Atomic Line Database (VALD)}, operated at Uppsala University, the Institute of Astronomy RAS in Moscow, and the University of Vienna.
    We acknowledge the use of the \href{http://simbad.u-strasbg.fr/simbad/}{SIMBAD database}. 
    This research has made use of the VizieR catalogue access tool, CDS, Strasbourg, France (\href{https://www.doi.org/10.26093/cds/vizier}{DOI:\@ 10.26093/cds/vizier}). The original description of the VizieR service was published in~\defcitealias{2000A&AS..143...23O}{A\&AS~143,~23}\citetalias{2000A&AS..143...23O}.
    This research has made use of NASA's \href{https://ui.adsabs.harvard.edu/}{Astrophysics Data System}.
    This work was carried out using the \swmf{} tools developed at The University of Michigan \href{https://spaceweather.engin.umich.edu/the-center-for-space-environment-modelling-csem/}{Center for Space Environment Modelling (CSEM)} and made available through the NASA \href{https://ccmc.gsfc.nasa.gov/}{Community Coordinated Modelling Center (CCMC)}.

    This work has made use of the following additional numerical software, statistics software and visualisation software: 
    NumPy version~1.19.4~\citep[][]{2011CSE....13b..22V},
    SciPy version~1.5.3~\citep[][]{2020SciPy-NMeth},
    Matplotlib version~3.3.3~\citep[][]{2007CSE.....9...90H},
    statsmodels version~0.12.1~\citep[][]{seabold2010statsmodels},
    Tecplot version~2020.2.0.110596, and PyTecplot version~1.3.3.

We would also like to thank the anonymous referee for their prompt response and diligence in helping to improve this manuscript.

\section*{Data availability}
The data underlying this article will be shared on reasonable request to the corresponding author.

%% file: tables/aggregate-magnetic-quantities.tex
\begin{tabular}{
        lSSSSSSS
        S[table-figures-decimal=0]
        S[table-figures-decimal=0]
    }
    \toprule
    Case & {\(|B_r|\)} & {\(\max |{B_r}|\)}  & {\(|\vec B|\)} & {Dip.} & {Quad.} & {Oct.} & {16+} & {\(\ell_{.90}\)} & {\(\ell_{.99}\)}\\
     & {\((\si{\gauss}) \)}
     & {\((\si{\gauss}) \)}
     & {\((\si{\gauss}) \)}
     & {\((\si{\percent}) \)}
     & {\((\si{\percent}) \)}
     & {\((\si{\percent}) \)}
     & {\((\si{\percent}) \)} \\
\midrule
\Unscaled{5}{C1}AV 523                   & 10.0649    & 36.7815    & 14.62      & 35.33      & 19.46      & 33.64      & 11.57      & 4          & 4         \\
\Unscaled{6}{C1}AV 1693                  & 20.4111    & 68.3811    & 28.98      & 32.72      & 46.88      & 14.77      & 5.63       & 3          & 6         \\
\Unscaled{7}{C1}AV 1826                  & 12.6199    & 49.2776    & 18.64      & 27.05      & 39.94      & 24.21      & 8.81       & 3          & 7         \\
\Unscaled{8}{C1}AV 2177                  & 5.5074     & 23.6902    & 8.72       & 54.09      & 24.91      & 7.49       & 13.52      & 4          & 6         \\
\Unscaled{9}{C1}TYC 1987                 & 14.6684    & 51.3897    & 21.40      & 35.13      & 24.18      & 13.30      & 27.39      & 5          & 5         \\
\midrule
\Scaled{5}{C1}$5\times$AV 523          & 50.2692    & 183.7643   & 71.48      & 35.63      & 19.73      & 33.30      & 11.35      & 4          & 4         \\
\Scaled{6}{C1}$5\times$AV 1693         & 101.9791   & 341.7769   & 143.63     & 32.91      & 46.76      & 14.76      & 5.57       & 3          & 6         \\
\Scaled{7}{C1}$5\times$AV 1826         & 63.0468    & 246.2808   & 91.19      & 27.29      & 39.69      & 24.29      & 8.73       & 3          & 7         \\
\Scaled{8}{C1}$5\times$AV 2177         & 27.5026    & 118.3420   & 40.60      & 54.56      & 24.41      & 7.75       & 13.28      & 4          & 6         \\
\Scaled{9}{C1}$5\times$TYC 1987        & 73.2733    & 256.8207   & 105.31     & 35.27      & 24.16      & 13.43      & 27.14      & 5          & 5         \\
\midrule
\Unscaled{A}{C2}DX Leo                   & 21.0633    & 73.2408    & 30.03      & 67.12      & 18.11      & 5.80       & 8.97       & 3          & 5         \\
\Unscaled{B}{C2}EP Eri                   & 9.5205     & 27.5817    & 14.16      & 29.39      & 63.16      & 0.92       & 6.53       & 2          & 5         \\
\Unscaled{C}{C2}HH Leo                   & 15.3462    & 62.4156    & 23.01      & 52.22      & 20.68      & 7.76       & 19.35      & 4          & 8         \\
\Unscaled{D}{C2}V439 And                 & 8.7439     & 35.1656    & 12.68      & 73.76      & 14.44      & 5.76       & 6.04       & 3          & 5         \\
\Unscaled{E}{C2}V447 Lac                 & 10.3606    & 60.6570    & 15.69      & 19.45      & 39.30      & 28.00      & 13.25      & 4          & 5         \\
\midrule
\Scaled{A}{C2}$5\times$DX Leo          & 105.2460   & 366.1075   & 148.06     & 66.66      & 18.55      & 5.69       & 9.10       & 3          & 5         \\
\Scaled{B}{C2}$5\times$EP Eri          & 47.5453    & 137.7644   & 68.56      & 29.93      & 62.68      & 0.90       & 6.50       & 2          & 5         \\
\Scaled{C}{C2}$5\times$HH Leo          & 76.6750    & 311.9413   & 112.91     & 52.18      & 20.79      & 7.66       & 19.37      & 4          & 8         \\
\Scaled{D}{C2}$5\times$V439 And        & 43.6575    & 175.6863   & 61.74      & 73.79      & 14.16      & 6.15       & 5.90       & 3          & 5         \\
\Scaled{E}{C2}$5\times$V447 Lac        & 51.7475    & 303.1658   & 76.28      & 19.31      & 39.27      & 28.03      & 13.39      & 4          & 5         \\
\bottomrule 
\end{tabular}

%% file: tables/main-table-body.tex
\begin{tabular}{
    l
    S  
    S[table-figures-exponent=0, table-figures-decimal=2, round-precision=2]  
    S[table-figures-exponent=0, table-figures-decimal=2, round-precision=2]  
    S[table-figures-exponent=0, table-figures-integer=2]  
    S[table-figures-exponent=0, table-figures-decimal=2, round-precision=2]  
    S[table-figures-exponent=0, table-figures-integer=2]  
    S[table-figures-exponent=0, table-figures-integer=2]  
    S  
    S  
    S  
    S[table-figures-exponent=0, table-figures-integer=2]  
}

\toprule
Case 
& {\(\SF\)}
& {\(\Phi_\text{open}\)}
& {\(S_\text{open}\)}
& {\(i_{B_r=0}\)}
& {\(\Phi_\text{axi}\)}
& {\(R_\Alfven\) }
& {\(\left|\vec r_\Alfven\times \uvec \Omega\right|\)}
& {\(\dot M\) }
& {\(\dot J\) }
& {\(P^\Earth_\Wind\)}
& {\(R_\text{m}\) }
\\ 
& {\(\left(\si{\weber}\right)\) } 
& {\(\left(\SF\right)\)}
& {\(\left(S\right)\)}
& {\(\left(\si{\degree}\right)\)}
& {\(\left(\Phi_\text{open}\right)\)}
& {\(\left(R_\Star{}\right)\)}
& {\(\left(R_\Star{}\right)\)}
& {\(\left(\si{\kilogram\per\second}\right)\)}
& {\(\left(\si{\newton\meter}\right)\)}
& {\(\left(\si{\pascal}\right)\)}
& {\(\left(R_\text{p}\right)\)}
\\ 
\midrule
\Unscaled{5}{C1}AV 523                   & 3.2e+15    & 0.27       & 0.23       & 3.9        & 0.99   & 15.3       & 11.5       & 1.9e+09    & 3.1e+23    & 7.6e-09    & 8.0       \\
\Unscaled{6}{C1}AV 1693                  & 8.6e+15    & 0.22       & 0.20       & 46.0       & 0.56   & 18.8       & 14.8       & 4.4e+09    & 1.8e+24    & 1.1e-08    & 7.5       \\
\Unscaled{7}{C1}AV 1826                  & 4.9e+15    & 0.24       & 0.15       & 14.9       & 0.89   & 15.3       & 11.7       & 2.9e+09    & 7.0e+23    & 8.1e-09    & 7.9       \\
\Unscaled{8}{C1}AV 2177                  & 2.0e+15    & 0.36       & 0.16       & 83.5       & 0.10   & 13.3       & 10.6       & 1.3e+09    & 2.8e+23    & 2.5e-09    & 9.6       \\
\Unscaled{9}{C1}TYC 1987                 & 6.1e+15    & 0.25       & 0.17       & 33.2       & 0.70   & 17.6       & 13.5       & 3.4e+09    & 1.1e+24    & 1.1e-08    & 7.5       \\ \midrule
\Scaled{5}{C1}$5\times$AV 523            & 1.6e+16    & 0.19       & 0.15       & 4.2        & 0.99   & 27.8       & 20.7       & 5.1e+09    & 2.1e+24    & 3.9e-08    & 6.1       \\
\Scaled{6}{C1}$5\times$AV 1693           & 4.3e+16    & 0.14       & 0.14       & 47.2       & 0.52   & 35.0       & 27.6       & 7.4e+09    & 8.2e+24    & 2.4e-08    & 6.6       \\
\Scaled{7}{C1}$5\times$AV 1826           & 2.5e+16    & 0.16       & 0.10       & 17.7       & 0.86   & 29.5       & 22.5       & 5.5e+09    & 3.8e+24    & 3.0e-08    & 6.3       \\
\Scaled{8}{C1}$5\times$AV 2177           & 1.0e+16    & 0.26       & 0.10       & 83.9       & 0.09   & 27.2       & 21.8       & 3.3e+09    & 2.2e+24    & 8.9e-09    & 7.8       \\
\Scaled{9}{C1}$5\times$TYC 1987          & 3.0e+16    & 0.16       & 0.12       & 33.0       & 0.68   & 31.7       & 24.5       & 5.9e+09    & 5.3e+24    & 2.6e-08    & 6.5       \\ \midrule
\Unscaled{A}{C2}DX Leo                   & 8.5e+15    & 0.28       & 0.17       & 80.7       & 0.13   & 25.6       & 20.5       & 3.8e+09    & 4.2e+24    & 6.3e-09    & 8.2       \\
\Unscaled{B}{C2}EP Eri                   & 3.0e+15    & 0.24       & 0.14       & 76.6       & 0.31   & 12.8       & 10.1       & 2.1e+09    & 4.8e+23    & 3.5e-09    & 9.1       \\
\Unscaled{C}{C2}HH Leo                   & 6.6e+15    & 0.29       & 0.13       & 80.3       & 0.15   & 21.6       & 17.4       & 3.2e+09    & 2.5e+24    & 5.3e-09    & 8.5       \\
\Unscaled{D}{C2}V439 And                 & 4.5e+15    & 0.35       & 0.21       & 9.8        & 0.96   & 16.3       & 12.3       & 3.3e+09    & 1.9e+24    & 1.5e-08    & 7.2       \\
\Unscaled{E}{C2}V447 Lac                 & 4.2e+15    & 0.25       & 0.23       & 26.7       & 0.77   & 15.1       & 11.5       & 2.3e+09    & 1.1e+24    & 6.3e-09    & 8.2       \\ \midrule
\Scaled{A}{C2}$5\times$DX Leo            & 4.2e+16    & 0.19       & 0.12       & 81.1       & 0.11   & 46.9       & 37.2       & 6.2e+09    & 2.0e+25    & 2.6e-08    & 6.5       \\
\Scaled{B}{C2}$5\times$EP Eri            & 1.5e+16    & 0.16       & 0.10       & 72.3       & 0.28   & 23.6       & 18.6       & 4.7e+09    & 2.9e+24    & 1.6e-08    & 7.0       \\
\Scaled{C}{C2}$5\times$HH Leo            & 3.3e+16    & 0.20       & 0.08       & 80.3       & 0.13   & 39.9       & 31.9       & 6.4e+09    & 1.6e+25    & 2.3e-08    & 6.6       \\
\Scaled{D}{C2}$5\times$V439 And          & 2.2e+16    & 0.24       & 0.13       & 10.2       & 0.95   & 30.3       & 22.5       & 7.1e+09    & 9.7e+24    & 5.6e-08    & 5.7       \\
\Scaled{E}{C2}$5\times$V447 Lac          & 2.1e+16    & 0.16       & 0.16       & 28.5       & 0.74   & 26.4       & 20.3       & 5.5e+09    & 7.2e+24    & 2.4e-08    & 6.6       \\
\bottomrule
\end{tabular}

%% file: tables/orthogonal_table.tex
\begin{tabular}{
    l
    S[
        table-figures-decimal=3, 
        table-figures-integer=1,  
        table-figures-uncertainty=4,
        table-figures-exponent = 0
    ]
    S[
        table-figures-decimal=1,
        table-figures-integer=3,
        table-figures-uncertainty= 0,
        table-figures-exponent = 0,
        table-space-text-post=$\, \%$,
    ]
    S[
        table-figures-decimal=3,
        table-figures-integer=1,
        table-figures-uncertainty= 0,
        table-figures-exponent = 0,
    ]
}
\toprule
{Quantity} & \multicolumn{3}{c}{Correlation with $\log_{10}\Phi$} \\
\cmidrule(lr){2-4}
{} & {$\displaystyle\alpha$} & {$\frac{y_\text{max}}{y_\text{min}}$} & {$r^2$} \\

\midrule
$\log_{10}|B_r|$ & 1.000+-0.000 & 162.2 \, \% & 0.976 \\
$\log_{10}\max|B_r|$ & 1.000+-0.000 & 186.4 \, \% & 0.937 \\
$\log_{10}|\vec{B}|$ & 0.983+-0.011 & 165.3 \, \% & 0.973 \\
$\log_{10} R_\Alfven$ & 0.394+-0.031 & 164.2 \, \% & 0.907 \\
$\log_{10} |\vec{r_\Alfven} \times \uvec{\Omega}|$ & 0.394+-0.032 & 171.0 \, \% & 0.877 \\
$\log_{10}\Phi_\text{open}$ & 0.745+-0.023 & 168.4 \, \% & 0.945 \\
$\log_{10}\dot M$ & 0.439+-0.090 & 203.5 \, \% & 0.883 \\
$\log_{10}\dot J$ & 1.094+-0.118 & 568.2 \, \% & 0.871 \\
$\log_{10}(\dot J/\Omega)$ & 1.094+-0.118 & 310.7 \, \% & 0.913 \\
$\log_{10}P_\Wind^\Earth$ & 0.736+-0.162 & 479.3 \, \% & 0.693 \\
$\log_{10} R_\text{m}$ & -0.123+-0.027 & 129.8 \, \% & 0.693 \\
\bottomrule
\end{tabular}

%% file: tables/correlations-rvalues-pvalues-predint-1-5.tex
\begin{tabular}{
    l
    S[table-figures-integer=2]
    S[
        table-figures-decimal=3,
        table-figures-integer=1,
        table-figures-uncertainty=0,
        table-figures-exponent = 0
    ]
    S[
        table-figures-decimal=2,
        table-figures-integer=1,
        table-figures-uncertainty=0,
        table-figures-exponent = 0
    ]
    S[
        table-figures-decimal=1,
        table-figures-integer=1,
        table-figures-uncertainty=0,
        table-figures-exponent = 3
    ]
    S[table-figures-integer=2]
    S[
        table-figures-decimal=3,
        table-figures-integer=1,
        table-figures-uncertainty=0,
        table-figures-exponent = 0
    ]
    S[
        table-figures-decimal=2,
        table-figures-integer=1,
        table-figures-uncertainty=0,
        table-figures-exponent = 0
    ]
    S[
        table-figures-decimal=1,
        table-figures-integer=1,
        table-figures-uncertainty=0,
        table-figures-exponent = 3
    ]
    }
    \toprule
{Quantity} & \multicolumn{8}{c}{Correlation with $a \log_{10}|B_r| + b$} \\
\cmidrule(lr){2-9}
    {} & \multicolumn{4}{c}{$B_\ZDI$ series}& \multicolumn{4}{c}{$5B_\ZDI$ series}\\
    \cmidrule(lr){2-5}\cmidrule(lr){6-9}
    {} & 
    {$a$} & {$r^2$} & {$\frac{y_{0.975}}{y_{0.025}}$} & {$p$} &
    {$a$} & {$r^2$} & {$\frac{y_{0.975}}{y_{0.025}}$} & {$p$} \\
    \midrule
$\log_{10}\max|B_r|$                                 &   0.94 \pm 0.25 & 0.836 &  2.34 &1.8e-06&   0.94 \pm 0.25 & 0.836 &  2.34 &1.8e-06\\
$\log_{10}|\vec{B}|$                                 &   0.95 \pm 0.03 & 0.997 &  1.12 &1.5e-17&   0.99 \pm 0.03 & 0.998 &  1.10 &1.3e-18\\
$\log_{10}\Phi$                                      &   1.01 \pm 0.20 & 0.900 &  1.99 &7.4e-08&   1.01 \pm 0.20 & 0.900 &  1.99 &7.3e-08\\
$\log_{10}\Phi_\text{open}$                          &   0.69 \pm 0.36 & 0.572 &  3.37 &1.1e-03&   0.65 \pm 0.35 & 0.554 &  3.30 &1.5e-03\\
$\log_{10} R_\Alfven$                                &   0.39 \pm 0.15 & 0.705 &  1.67 &9.1e-05&   0.30 \pm 0.17 & 0.525 &  1.80 &2.2e-03\\
$\log_{10} |\vec{r_\Alfven} \times \uvec{\Omega}|$   &   0.38 \pm 0.17 & 0.648 &  1.78 &2.9e-04&   0.30 \pm 0.19 & 0.465 &  1.92 &5.1e-03\\
$\log_{10}\dot M$                                    &   0.56 \pm 0.31 & 0.538 &  2.91 &1.9e-03&   0.31 \pm 0.23 & 0.396 &  2.20 &1.2e-02\\
$\log_{10}\dot J$                                    &   1.39 \pm 0.76 & 0.546 & 13.26 &1.6e-03&   1.02 \pm 0.71 & 0.421 & 11.42 &8.9e-03\\
$\log_{10}\dot J / \Omega$                           &   1.22 \pm 0.67 & 0.543 &  9.74 &1.7e-03&   0.84 \pm 0.60 & 0.414 &  7.74 &9.7e-03\\
$\log_{10}P_\Wind^\Earth$                      &   0.64 \pm 0.55 & 0.328 &  6.52 &2.6e-02&   0.59 \pm 0.58 & 0.269 &  7.24 &4.7e-02\\
$\log_{10} R_\text{m}$                             &  -0.11 \pm 0.09 & 0.328 &  1.37 &2.6e-02&  -0.10 \pm 0.10 & 0.269 &  1.39 &4.7e-02\\
\midrule
{Quantity} & \multicolumn{8}{c}{Correlation with $a \log_{10}\Phi + b$} \\
\cmidrule(lr){2-9}
    {} & \multicolumn{4}{c}{$B_\ZDI$ series}& \multicolumn{4}{c}{$5B_\ZDI$ series}\\
    \cmidrule(lr){2-5}\cmidrule(lr){6-9}
    {} & 
    {$a$} & {$r^2$} & {$\frac{y_{0.975}}{y_{0.025}}$} & {$p$} &
    {$a$} & {$r^2$} & {$\frac{y_{0.975}}{y_{0.025}}$} & {$p$} \\
    \midrule
$\log_{10}|B_r|$                                     &   0.89 \pm 0.18 & 0.900 &  1.91 &7.4e-08&   0.89 \pm 0.18 & 0.900 &  1.91 &7.3e-08\\
$\log_{10}\max|B_r|$                                 &   0.87 \pm 0.25 & 0.816 &  2.46 &4.0e-06&   0.87 \pm 0.25 & 0.816 &  2.46 &3.9e-06\\
$\log_{10}|\vec{B}|$                                 &   0.84 \pm 0.18 & 0.884 &  1.94 &1.9e-07&   0.88 \pm 0.18 & 0.891 &  1.95 &1.3e-07\\
$\log_{10}\Phi_\text{open}$                          &   0.78 \pm 0.21 & 0.825 &  2.18 &2.9e-06&   0.73 \pm 0.22 & 0.800 &  2.22 &6.8e-06\\
$\log_{10} R_\Alfven$                                &   0.38 \pm 0.13 & 0.769 &  1.58 &1.8e-05&   0.31 \pm 0.15 & 0.616 &  1.70 &5.3e-04\\
$\log_{10} |\vec{r_\Alfven} \times \uvec{\Omega}|$   &   0.38 \pm 0.14 & 0.715 &  1.68 &7.2e-05&   0.31 \pm 0.16 & 0.552 &  1.82 &1.5e-03\\
$\log_{10}\dot M$                                    &   0.64 \pm 0.20 & 0.788 &  2.06 &1.0e-05&   0.37 \pm 0.17 & 0.640 &  1.84 &3.4e-04\\
$\log_{10}\dot J$                                    &   1.55 \pm 0.50 & 0.773 &  6.22 &1.6e-05&   1.17 \pm 0.53 & 0.637 &  6.88 &3.6e-04\\
$\log_{10}\dot J / \Omega$                           &   1.39 \pm 0.41 & 0.806 &  4.40 &5.5e-06&   1.01 \pm 0.42 & 0.679 &  4.54 &1.6e-04\\
$\log_{10}P_\Wind^\Earth$                      &   0.72 \pm 0.46 & 0.472 &  5.27 &4.6e-03&   0.58 \pm 0.53 & 0.299 &  6.96 &3.5e-02\\
$\log_{10} R_\text{m}$                             &  -0.12 \pm 0.08 & 0.472 &  1.32 &4.6e-03&  -0.10 \pm 0.09 & 0.299 &  1.38 &3.5e-02\\
\midrule
{Quantity} & \multicolumn{8}{c}{Correlation with $a \log_{10}\Phi_\text{open} + b$} \\
\cmidrule(lr){2-9}
    {} & \multicolumn{4}{c}{$B_\ZDI$ series}& \multicolumn{4}{c}{$5B_\ZDI$ series}\\
    \cmidrule(lr){2-5}\cmidrule(lr){6-9}
    {} & 
    {$a$} & {$r^2$} & {$\frac{y_{0.975}}{y_{0.025}}$} & {$p$} &
    {$a$} & {$r^2$} & {$\frac{y_{0.975}}{y_{0.025}}$} & {$p$} \\
    \midrule
$\log_{10}|B_r|$                                     &   0.83 \pm 0.43 & 0.572 &  3.80 &1.1e-03&   0.85 \pm 0.46 & 0.554 &  3.91 &1.5e-03\\
$\log_{10}\max|B_r|$                                 &   0.82 \pm 0.46 & 0.532 &  4.19 &2.0e-03&   0.85 \pm 0.49 & 0.524 &  4.25 &2.3e-03\\
$\log_{10}|\vec{B}|$                                 &   0.78 \pm 0.42 & 0.549 &  3.70 &1.6e-03&   0.83 \pm 0.46 & 0.538 &  3.97 &1.9e-03\\
$\log_{10}\Phi$                                      &   1.06 \pm 0.29 & 0.825 &  2.49 &2.9e-06&   1.09 \pm 0.33 & 0.800 &  2.65 &6.8e-06\\
$\log_{10} R_\Alfven$                                &   0.46 \pm 0.13 & 0.805 &  1.52 &5.7e-06&   0.43 \pm 0.12 & 0.812 &  1.45 &4.6e-06\\
$\log_{10} |\vec{r_\Alfven} \times \uvec{\Omega}|$   &   0.47 \pm 0.14 & 0.791 &  1.56 &9.3e-06&   0.44 \pm 0.14 & 0.767 &  1.54 &1.9e-05\\
$\log_{10}\dot M$                                    &   0.78 \pm 0.20 & 0.845 &  1.85 &1.3e-06&   0.46 \pm 0.20 & 0.667 &  1.80 &2.0e-04\\
$\log_{10}\dot J$                                    &   1.92 \pm 0.46 & 0.863 &  4.14 &5.7e-07&   1.61 \pm 0.47 & 0.808 &  4.06 &5.2e-06\\
$\log_{10}\dot J / \Omega$                           &   1.80 \pm 0.13 & 0.987 &  1.48 &1.4e-13&   1.47 \pm 0.17 & 0.965 &  1.65 &7.6e-11\\
$\log_{10}P_\Wind^\Earth$                      &   0.79 \pm 0.57 & 0.408 &  5.81 &1.0e-02&   0.54 \pm 0.71 & 0.174 &  8.21 &1.2e-01\\
$\log_{10} R_\text{m}$                             &  -0.13 \pm 0.09 & 0.408 &  1.34 &1.0e-02&  -0.09 \pm 0.12 & 0.174 &  1.42 &1.2e-01\\
\bottomrule
\end{tabular}

%% file: pooled-series.tex
\section{Pooled series}\label{sec:pooled-series}
\begin{figure*}
    \centering
    \includegraphics{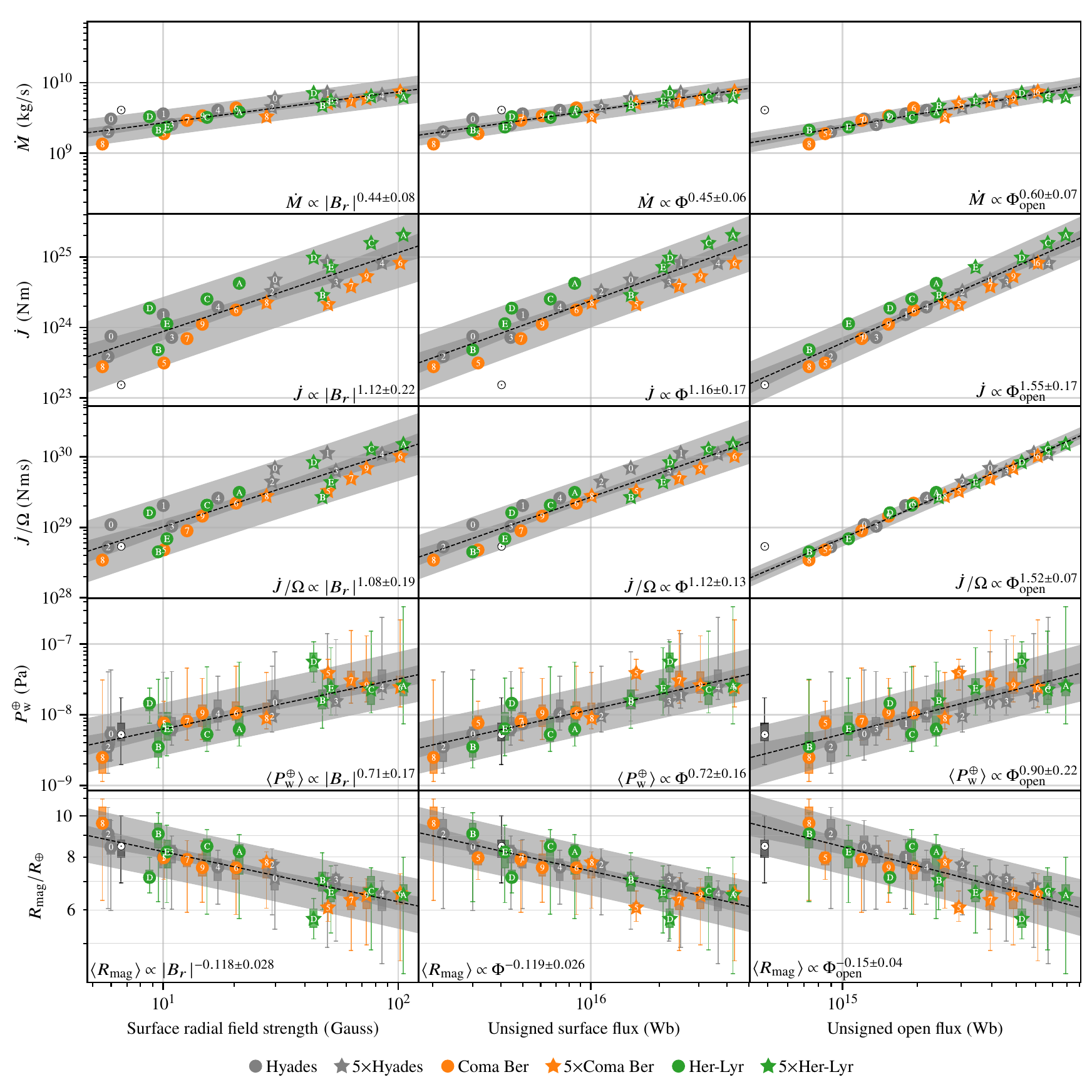}
    \caption{
        Trend lines and variation bands as in Fig.~\ref{fig:trend-5x3-broken}, but treating the \(B_\ZDI\) and \(5B_\ZDI\) series as a single population. The fitted curve is also described in Table~\ref{tab:correlations-pooled}.
    }\label{fig:trend-5x3} 
\end{figure*}
\begin{table*}
    \centering
    \caption{
        This table is similar to 
        Table~\ref{tab:correlations-split}, 
        but for the pooled series comprising all the models of both the \(B_\ZDI\) and \(5B_\ZDI\) series. As the fitted curves here maximise the coefficients of determination, the  \(r^2\) values are higher than the \(r^2\) values of the geometric midpoint approach in Table~\ref{tab:orthogonal_correlations}. 
    }\label{tab:correlations-pooled}
    \sisetup{
        table-figures-decimal=2,
        table-figures-integer=1,
        table-figures-uncertainty=2,
        table-number-alignment=center,
        add-decimal-zero=false,
        add-integer-zero = false,
        omit-uncertainty = false
    }
    \input{tables/correlations-rvalues-pvalues-predint-Full.tex}
\end{table*}

Here present the same analysis as in Section~\ref{sec:trends-and-correlations}, but for a pooled data series comprising the values in both the \(B_\ZDI\) and the \(5B_\ZDI\) series. The fitted curves are found in Fig.~\ref{fig:trend-5x3} and the corresponding data is found in Table~\ref{tab:correlations-pooled}. Pooling the data gives a single series of data with a larger range of \(|B_r|\) values and \(2\times15=30\) data points. 

In \citetalias{paper1} we observed the that the correlations in the \(|B_r|\) improved and seemed to approach the correlations in the \(\Phi_\text{open}\) series. The addition of the data points of this work show that this effect was largely spurious except for the wind pressure for an Earth-like planet and the magnetospheric stand-off distance. The other parameters \(\dot M\), \(\dot J\), and the new parameter \(\dot J/\Omega\) have a significantly tighter correlation  with \(\Phi_\text{open}\) than with \(|B_r|\), with \(\Phi\) occupying a middle ground. 

When considering the pooled series and its large \(B_r\) range it should be kept in mind that here, the rotation-field strength correlation \citep{1984ApJ...279..763N,2018haex.bookE..26V} is adjusted for. This is different from studies of stars over a large age range, such as \citet{2015MNRAS.449.4117V} and \citet{2018ApJ...856...53P}, where this correlation, being one of the objects of study, is intentionally present in the final results.

%% file: tables/correlations-rvalues-pvalues-predint-Full.tex
\begin{tabular}{
    l
    S[table-figures-integer=2]
    S[
        table-figures-decimal=3,
        table-figures-integer=1,
        table-figures-uncertainty=0,
        table-figures-exponent = 0
    ]
    S[
        table-figures-decimal=2,
        table-figures-integer=1,
        table-figures-uncertainty=0,
        table-figures-exponent = 0
    ]
    S[
        table-figures-decimal=1,
        table-figures-integer=1,
        table-figures-uncertainty=0,
        table-figures-exponent = 3
    ]
    }
    \toprule
{Quantity} & \multicolumn{4}{c}{Correlation with $a \log_{10}|B_r| + b$} \\

    \cmidrule(lr){2-5}
    {} & 
    {$a$} & {$r^2$} & {$\frac{y_{0.975}}{y_{0.025}}$} & {$p$} \\
    \midrule
$\log_{10}\max|B_r|$                                 &   0.99 \pm 0.07 & 0.964 &  2.14 &1.0e-21\\
$\log_{10}|\vec{B}|$                                 &   0.98 \pm 0.01 & 0.999 &  1.11 &5.0e-46\\
$\log_{10}\Phi$                                      &   1.00 \pm 0.06 & 0.977 &  1.84 &1.9e-24\\
$\log_{10}\Phi_\text{open}$                          &   0.73 \pm 0.11 & 0.878 &  2.93 &2.5e-14\\
$\log_{10} R_\Alfven$                                &   0.38 \pm 0.05 & 0.902 &  1.65 &1.3e-15\\
$\log_{10} |\vec{r_\Alfven} \times \uvec{\Omega}|$   &   0.38 \pm 0.05 & 0.881 &  1.75 &1.9e-14\\
$\log_{10}\dot M$                                    &   0.44 \pm 0.08 & 0.802 &  2.37 &2.3e-11\\
$\log_{10}\dot J$                                    &   1.12 \pm 0.22 & 0.793 &  9.55 &4.3e-11\\
$\log_{10}\dot J / \Omega$                           &   1.08 \pm 0.19 & 0.828 &  7.03 &3.3e-12\\
$\log_{10}P_\Wind^\Earth$                      &   0.71 \pm 0.17 & 0.728 &  5.58 &2.1e-09\\
$\log_{10} R_\text{m}$                             &  -0.12 \pm 0.03 & 0.728 &  1.33 &2.1e-09\\
\midrule
{Quantity} & \multicolumn{4}{c}{Correlation with $a \log_{10}\Phi + b$} \\

    \cmidrule(lr){2-5}
    {} & 
    {$a$} & {$r^2$} & {$\frac{y_{0.975}}{y_{0.025}}$} & {$p$} \\
    \midrule
$\log_{10}|B_r|$                                     &   0.97 \pm 0.06 & 0.977 &  1.83 &1.9e-24\\
$\log_{10}\max|B_r|$                                 &   0.97 \pm 0.08 & 0.957 &  2.29 &1.2e-20\\
$\log_{10}|\vec{B}|$                                 &   0.95 \pm 0.06 & 0.974 &  1.87 &1.0e-23\\
$\log_{10}\Phi_\text{open}$                          &   0.75 \pm 0.07 & 0.948 &  2.01 &1.5e-19\\
$\log_{10} R_\Alfven$                                &   0.38 \pm 0.04 & 0.921 &  1.57 &6.2e-17\\
$\log_{10} |\vec{r_\Alfven} \times \uvec{\Omega}|$   &   0.38 \pm 0.05 & 0.901 &  1.66 &1.4e-15\\
$\log_{10}\dot M$                                    &   0.45 \pm 0.06 & 0.885 &  1.93 &1.1e-14\\
$\log_{10}\dot J$                                    &   1.16 \pm 0.17 & 0.874 &  5.83 &4.1e-14\\
$\log_{10}\dot J / \Omega$                           &   1.12 \pm 0.13 & 0.913 &  4.01 &2.3e-16\\
$\log_{10}P_\Wind^\Earth$                      &   0.72 \pm 0.16 & 0.761 &  5.02 &3.4e-10\\
$\log_{10} R_\text{m}$                             &  -0.12 \pm 0.03 & 0.761 &  1.31 &3.4e-10\\
\midrule
{Quantity} & \multicolumn{4}{c}{Correlation with $a \log_{10}\Phi_\text{open} + b$} \\

    \cmidrule(lr){2-5}
    {} & 
    {$a$} & {$r^2$} & {$\frac{y_{0.975}}{y_{0.025}}$} & {$p$} \\
    \midrule
$\log_{10}|B_r|$                                     &   1.20 \pm 0.17 & 0.878 &  3.98 &2.5e-14\\
$\log_{10}\max|B_r|$                                 &   1.20 \pm 0.18 & 0.867 &  4.28 &9.0e-14\\
$\log_{10}|\vec{B}|$                                 &   1.18 \pm 0.17 & 0.873 &  4.00 &4.8e-14\\
$\log_{10}\Phi$                                      &   1.27 \pm 0.11 & 0.948 &  2.49 &1.5e-19\\
$\log_{10} R_\Alfven$                                &   0.51 \pm 0.05 & 0.947 &  1.45 &2.2e-19\\
$\log_{10} |\vec{r_\Alfven} \times \uvec{\Omega}|$   &   0.51 \pm 0.05 & 0.938 &  1.50 &2.0e-18\\
$\log_{10}\dot M$                                    &   0.60 \pm 0.07 & 0.906 &  1.81 &6.1e-16\\
$\log_{10}\dot J$                                    &   1.55 \pm 0.17 & 0.927 &  3.82 &1.9e-17\\
$\log_{10}\dot J / \Omega$                           &   1.52 \pm 0.07 & 0.987 &  1.71 &5.3e-28\\
$\log_{10}P_\Wind^\Earth$                      &   0.90 \pm 0.22 & 0.707 &  5.96 &6.0e-09\\
$\log_{10} R_\text{m}$                             &  -0.15 \pm 0.04 & 0.707 &  1.35 &6.0e-09\\
\bottomrule
\end{tabular}